\documentclass[10pt,journal]{IEEEtran}
\ifCLASSOPTIONcompsoc
  \usepackage[nocompress]{cite}
\else
  \usepackage{cite}
\fi

\usepackage{amsmath,amssymb,amsfonts}
\usepackage{algorithm}  
\usepackage{algpseudocode}
\usepackage{graphicx}
\usepackage{textcomp}
\usepackage{xcolor}
\usepackage{bm}
\usepackage{booktabs}
\usepackage{array}
\usepackage{diagbox} 
\usepackage{caption}
\usepackage{colortbl}
\usepackage{tabu}
\usepackage{subfigure}
\usepackage{soul}
\usepackage[numbers,sort&compress]{natbib}
\usepackage{lipsum}
\usepackage{mathtools}
\usepackage{cuted}
\usepackage{ragged2e}
\usepackage[super]{nth}
\ifCLASSINFOpdf
\else
\fi

\begin{document}
\title{Multi-Agent Deep Reinforcement Learning Multiple Access for Heterogeneous Wireless Networks with Imperfect Channels}
\author{Yiding Yu,~\IEEEmembership{Student Member,~IEEE,}
        Soung Chang Liew,~\IEEEmembership{Fellow,~IEEE,} \\
        and Taotao Wang,~\IEEEmembership{Member,~IEEE}
\IEEEcompsocitemizethanks{
\IEEEcompsocthanksitem Y. Yu and S.C. Liew are with the Department of Information Engineering, The Chinese University of Hong Kong, Shatin, Hong Kong Special Administrative Region, China. Email: \{yy016, soung\}@ie.cuhk.edu.hk 
\IEEEcompsocthanksitem T. Wang was with the Department of Information Engineering, The Chinese University of Hong Kong, Shatin, Hong Kong Special Administrative Region, China. He is now with the College
of Information Engineering, Shenzhen University, Shenzhen, China. Email: ttwang@ie.cuhk.edu.hk
}
}

\IEEEtitleabstractindextext{%
\begin{abstract} 
    \justifying{This paper investigates a futuristic spectrum sharing paradigm for heterogeneous wireless networks with imperfect channels. In the heterogeneous networks, multiple wireless networks adopt different medium access control (MAC) protocols to share a common wireless spectrum and each network is unaware of the MACs of others. This paper aims to design a distributed deep reinforcement learning (DRL) based MAC protocol for a particular network, and the objective of this network is to achieve a global  $\alpha$-fairness objective. In the conventional DRL framework, feedback/reward given to the agent is always correctly received, so that the agent can optimize its strategy based on the received reward. In our wireless application where the channels are noisy, the feedback/reward (i.e., the ACK packet) may be lost due to channel noise and interference. Without correct feedback, the agent (i.e., the network user) may fail to find a good solution. Moreover, in the distributed protocol, each agent makes decisions on its own. It is a challenge to guarantee that the multiple agents will make coherent decisions and work together to achieve the same objective, particularly in the face of imperfect feedback channels. To tackle the challenge, we put forth (i) a feedback recovery mechanism to recover missing feedback information, and (ii) a two-stage action selection mechanism to aid coherent decision making to reduce transmission collisions among the agents. Extensive simulation results demonstrate the effectiveness of these two mechanisms. Last but not least, we believe that the feedback recovery mechanism and the two-stage action selection mechanism can also be used in general distributed multi-agent reinforcement learning problems in which feedback information on rewards can be corrupted.}
\end{abstract}
\begin{IEEEkeywords}
	MAC protocol, heterogeneous wireless networks, imperfect channel, multi-agent, deep reinforcement learning. 
\end{IEEEkeywords}}

\maketitle

\IEEEdisplaynontitleabstractindextext
\IEEEpeerreviewmaketitle

\section{Introduction}\label{sec:introduction}
	This paper investigates a futuristic spectrum sharing paradigm where multiple separate wireless networks adopt different media access control (MAC) protocols to transmit packets on a common wireless spectrum\footnote{This futuristic spectrum sharing paradigm, wherein heterogeneous networks collaborate to use the shared spectrum without detailed knowledge of each other's MAC, was first envisioned by the DARPA Spectrum Collaboration Challenge (SC2) competition \cite{tilghman2019will} to spur the development of next-generation wireless networks. Our current paper investigates a particular scenario in which the heterogeneous networks share the spectrum in the time domain.}. In sharing the spectrum, each network must respect spectrum usage by other networks and must not hog the spectrum to the detriment of other networks. Unlike in the conventional cognitive radio networks \cite{liang2008sensing}, all network users in our scenario are on an equal footing in that they are not divided into primary users and secondary users. This paper puts forth an intelligent MAC protocol for a particular network---among the networks sharing the spectrum---to achieve efficient and equitable spectrum sharing among all networks. A major challenge for the intelligent MAC protocol is that the particular network concerned does not know the MAC protocols used by other networks. Our MAC protocol must learn the medium-access behavior of other networks on the fly so that harmonious coexistence and fair sharing of spectrum with them can be achieved. 
	
	The fundamental technique in our MAC protocol design is deep reinforcement learning (DRL). DRL is a machine learning technique that combines the decision-making ability of reinforcement learning (RL) \cite{sutton2018reinforcement} and the function approximation ability of deep neural networks \cite{lecun2015deep} to solve complex decision-making problems, including game playing, robot control, wireless communications, and network management and control \cite{DQNpaper,silver2016mastering, gu2017deep,luong2019applications, sun2019application,shao2019significant}. In RL/DRL, in each time step, the decision-making agent interacts with its external environment by executing an action. The agent then receives feedback in the form of a reward that tells the agent how good the action was. The agent strives to optimize the rewards it receives over its lifetime \cite{sutton2018reinforcement}.
	
	A key advantage of our DRL based MAC protocol is that it can learn to coexist with other MAC protocols without knowing their operational details. In our paper, we refer to our DRL MAC protocol as deep reinforcement learning multiple access (DLMA), and network adopting DLMA as the DLMA network. Within the DLMA network, each user is regarded as a DRL agent and it employs the DLMA protocol to make its MAC decisions, i.e., to transmit or not to transmit data packets. 
	
	To enable efficient and equitable spectrum sharing among all the users, the agents adopt ``$\alpha$-fairness'' \cite{mo2000fair} as their objective. The  $\alpha$-fairness objective is global and general in that (i) each agent not only optimizes its own spectrum usage, but also optimizes the spectrum usage of other agents and the non-agent users of coexisting networks; (ii) the agents can achieve different specific objectives by adjusting the value of the parameter  $\alpha$, e.g.,  $\alpha = 0$ corresponds to maximizing the sum throughput of all users and  $\alpha = 1$ corresponds to achieving proportional fairness among all users.
	
	The conventional RL/DRL framework has inherent limitations. First, the conventional RL/DRL framework only aims to maximize the cumulative discounted rewards of the agent, i.e., the objective to be optimized by the agent is a weighted sum of the rewards. However, the  $\alpha$-fairness objective function, in general, is a nonlinear combination of utility functions of all users \cite{mo2000fair}. Therefore, in our design, we adopt the multi-dimensional DRL framework in \cite{yu2019deep} to solve this problem. 
	
	Second, in the conventional RL/DRL framework, the feedback to the agent about the reward is assumed to be always correctly received. However, in the wireless environment, the feedback may be lost due to noise and interference in the wireless channel. Without the correct reward values, the agent may fail to find the optimal strategy. This paper puts forth a \textit{feedback recovery mechanism} for incorporation into DLMA to reduce the detrimental effects of imperfect channels. The key idea is that we feedback the reward of the current time step not just in the current time step, but also in the next $K - 1$   time steps. Thus, an earlier missing reward may be recovered in a later time step. The essence is that late learning is better than no learning. 
	
	Third, the conventional RL/DRL framework only works for single-agent problems. For multi-agent problems where each agent makes decisions on its own without collaboration with other agents, it is difficult to guarantee that multiple agents will work together to achieve the same objective \cite{zhang2019multi}. To avoid conflicting decisions among the agents, we put forth a \textit{two-stage action selection mechanism} in DLMA. In this mechanism, each agent first decides on the ``network action'' for the DLMA network as to whether a DLMA agent should transmit a packet or not. The network action can be regarded as a collective policy of all the agents. If the decision is to transmit, an agent then decides as to whether it is the one that will transmit. With this mechanism,  the agents strive to coexist with the non-agent users of the other networks harmoniously in the first decision stage and reduce collisions among the agents in the second decision stage.
	
	Extensive simulation results show that our feedback recovery mechanism can effectively reduce the detrimental effects of the imperfect feedback channels in the heterogeneous wireless networks. For benchmarking, we replace the agents (i.e., the DLMA users) with model-aware users that are aware of the operational details of the other MACs of the coexisting networks and we assume the feedback channels of the model-aware users are perfect. We demonstrate that the results achieved by DLMA can approximate the optimal benchmarks even though DLMA is deployed in an imperfect channel setting. We also demonstrate the capability of our two-stage action selection mechanism in reducing collisions among the multiple agents. Specifically, we demonstrate that when the channels of the agents are ``perfect'' or ``imperfect but dependent'', collisions can be avoided among the agents; when the channels are ``imperfect and independent'', collisions among the agents cannot be eliminated but can be significantly reduced. 
	
	Overall, the main contributions of this paper are summarized as follows:
	\begin{enumerate}
		\item We develop a distributed DRL based MAC protocol for efficient and equitable spectrum sharing in heterogeneous wireless networks with imperfect channels. 
		\item We demonstrate that our proposed feedback recovery mechanism can effectively solve the problem of imperfect feedback channels. We also demonstrate collisions among the multiple distributed agents can be significantly reduced with our two-stage action selection mechanism. 
		\item We believe that the feedback recovery mechanism and the two-stage action selection mechanism in our MAC protocol design can also find use in a wide range of applications. Specifically, the idea of feedback/reward recovery can be applied in the general corruption reward problem \cite{ijcai2017-656, wang2020rlnoisy} in reinforcement learning and the two-stage action selection process can be used in the distributed multi-agent problem where the agents need to avoid conflicts with each other \cite{zhang2019multi}. 
	\end{enumerate}
	
	\subsection{Related Work}
	In our previous work \cite{yu2019deep}, we developed a DRL based MAC protocol for heterogeneous wireless networks with perfect channels. In \cite{yu2019deep}, the access point (AP) is regarded as a centralized DRL agent. The agent is responsible for making MAC decisions and it broadcasts the control information containing the decisions to the users, telling them who should transmit in what slots. However, for a practical scenario with noisy wireless channels, the control information may be lost.  Without immediate and correct control information in a particular time slot, the users will not know whether to transmit or not.  Thus, a method is needed so that users can make appropriate decisions even without immediate feedback. 
	
	The current paper provides a method to do so. Specifically, this paper removes the perfect channel assumption in \cite{yu2019deep} and puts forth a distributed DRL MAC protocol for heterogeneous wireless networks with imperfect channels. In this paper, rather than having the AP serving as the single centralized agent in the whole network,  each user is a DRL agent, and each agent makes its own MAC decisions. We put forth a two-stage action selection mechanism to reduce collisions among the agents' transmissions. When the channels are imperfect, the conventional DRL techniques with the implicit assumption of perfect feedback are not suitable anymore. Therefore, this paper proposes a feedback recovery mechanism to reduce the detrimental effects of imperfect feedback channels. The detailed differences, especially the difference in DRL formulations, between \cite{yu2019deep} and our current paper will be discussed in the main body of our paper.

	Apart from \cite{yu2019deep}, there have also been other work on DRL based MAC.  As in \cite{yu2019deep}, investigations in \cite{naparstek2018deep,wang2018deep,chang2018distributive,zhong2018actor,xu2018deep, tan2019deep} also assumed the physical channels are perfect and did not have a scheme to deal with imperfect channels. Other fine differences between \cite{naparstek2018deep,wang2018deep,chang2018distributive,zhong2018actor,xu2018deep,tan2019deep} and our current work are elaborated in the following. 
	
	The MAC in \cite{naparstek2018deep} is designed for homogeneous wireless networks, where all the users adopt the same MAC protocol to dynamically access multiple wireless channels.  By contrast, our MAC protocol is targeted for heterogeneous networks in which our MAC must learn to coexist with other MAC protocols. Both \cite{naparstek2018deep} and our current paper consider achieving a global  $\alpha$-fairness objective. However, the ``fairness'' in \cite{naparstek2018deep} is achieved among the homogeneous DRL users. By contrast, our DRL users aim to achieve ``fairness'' among the heterogeneous DRL users and non-DRL users. 
	
	The authors in \cite{wang2018deep,chang2018distributive,zhong2018actor,xu2018deep} also investigated the multi-channel access problem as in \cite{naparstek2018deep}. The difference is that in \cite{naparstek2018deep}, the authors assumed the channels are time-invariant, whereas the channels in \cite{wang2018deep,chang2018distributive,zhong2018actor,xu2018deep} may be time-varying in that some “primary” or ``legacy'' users may occupy the channels from time to time. Therefore, the wireless networks in \cite{wang2018deep,chang2018distributive,zhong2018actor,xu2018deep} can also be regarded as ``heterogeneous''. However, the DRL users in \cite{wang2018deep,chang2018distributive,zhong2018actor,xu2018deep} aim to maximize their own throughputs by learning the channel characteristics and the transmission patterns of the ``primary'' or ``legacy'' users. By contrast, the DRL users in our work (also in \cite{naparstek2018deep}) aim to achieve a global $\alpha$-fairness objective, which includes achieving maximum sum throughput, proportional fairness, and max-min fairness as subcases \cite{mo2000fair}.
	
	The authors in \cite{tan2019deep}  investigated heterogeneous wireless networks. Specifically, in \cite{tan2019deep}, an LTE network exercises a coarse-grained DRL based MAC control to coexist with a WiFi network. The LTE MAC in \cite{tan2019deep} decides a period for LTE transmissions and a period for WiFi transmissions. During the corresponding periods, LTE and WiFi transmit packets without interfering with each other. By contrast, the MAC of our design exercises fine-grained control in that our MAC makes decisions (i.e., to transmit or not to transmit) on a packet-by-packet basis. Furthermore, in \cite{tan2019deep}, the LTE network is model-aware in that the LTE network knows the coexisting network is WiFi. Therefore, the approach in \cite{tan2019deep} is not generalizable to situations where the LTE network coexists with other networks. By contrast, our DRL based MAC protocol is model-free in that it does not presume knowledge of the operational details of the MAC protocols of coexisting networks.

	\section{Overview of Reinforcement Learning}
	The underpinning technique in our DLMA algorithm is deep reinforcement learning, especially the Deep Q-Network (DQN) algorithm \cite{DQNpaper}. This section first presents Q-learning, a representative reinforcement learning (RL) algorithm. After that, the DQN algorithm is introduced. 
	
	\subsection{Q-learning}
\begin{figure}[!t]
		\centering
		\includegraphics[scale=0.9]{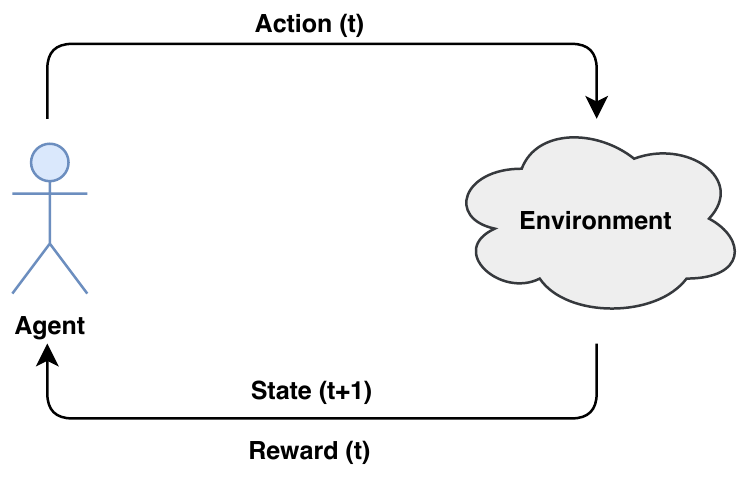}
		\caption[]{Reinforcement learning framework: the interaction between an agent and its environment.}
		\label{fig:RL_framework}
	\end{figure}

	In the RL framework, an agent interacts with an external environment in discrete time steps \cite{sutton2018reinforcement}, as illustrated in Fig. \ref{fig:RL_framework}. Particularly, in time step  $t$, given the environment state  $s_t$, the agent takes an action  $a_t$ according to a policy  $\pi $, which maps the states of the environment to the actions of the agent. After receiving action  $a_t$, the environment feeds back a reward  $r_t$ to the agent to evaluate the agent's performance in time step   $t$. In addition, the environment state transmits to $s_{t+1}$  in time step  $t+1$. The goal of the agent is to find an optimal policy ${\pi ^ * }$ that maximizes the cumulative discounted rewards  ${R_t} = \sum\nolimits_{k = 0}^\infty  {{\gamma ^k}  {r_{t + k}}}$, where  $\gamma  \in \left( {0,1} \right]$  is a discount factor. 
	
	For a particular policy $\pi$ and a particular state-action pair  $\left( {s,a} \right)$,  Q-learning captures the expected cumulative discounted rewards with an action-value function, i.e., the Q-function:  ${Q^\pi }\left( {s,a} \right) = {\bf{E}}\left[ {{R_t}|{s_t} = s,{a_t} = a,\pi } \right]$. The Q function of the optimal policy  ${\pi ^ * }$ is given by  ${Q^*}\left( {s,a} \right) = {\max _\pi }{Q^\pi }\left( {s,a} \right)$.  To find the optimal policy  ${\pi ^ * }$, the Q-learning agent maintains the Q function,  $Q(s,a)$, for any state-action pair   $\left( {s,a} \right)$,  in a tabular form. In time step  $t$,  given state  $s_t$, the agent selects an action ${a_t} = \arg {\max _a}Q({s_t},a)$  based on its current Q table. After receiving the reward  $r_t$  and observing the new state  $s_{t+1}$, the agent constructs an experience tuple  ${e_t} = ({s_t},{a_t},{r_t},{s_{t + 1}})$. Then the experience tuple  $e_t$ is used to update $Q({s_t},{a_t})$  as follows: 	
	\begin{align}
	\footnotesize
	Q\left( {{s_t},{a_t}} \right) \leftarrow &\left( {1 - \beta } \right)Q\left( {{s_t},{a_t}} \right) + \nonumber \\
	&\beta  \left( {{r_t} + \gamma   \mathop {\max }\limits_{a'} Q\left( {{s_{t + 1}},a'} \right) - Q\left( {{s_t},{a_t}} \right)} \right),
	\end{align}
	where  $\beta  \in (0,1]$ is the learning rate of the algorithm.
	
	In Q-learning, in addition to selecting action using  ${a_t} = \arg {\max _a}Q({s_t},a)$, the  $\epsilon$-greedy algorithm is often adopted. Specifically, the action ${a_t} = \arg {\max _a}Q({s_t},a)$  is chosen with probability  $1 - \epsilon$, and a random action is chosen uniformly among all  actions with probability  $\epsilon$.  The random action selection avoids the algorithm from zooming into a local optimal policy and allows the agent to explore a wider spectrum of different actions in search of the optimal policy. 
	
	One thing to be pointed out here is that Q-learning is a model-free algorithm in that it does not need to know the state transition probability. The Q-learning agent learns to find the optimal policy through the experiences obtained during the interaction with the environment. 
	
	\subsection{Deep Q-Network}\label{sec:DQN}
	In a stationary environment, Q-learning is shown to converge if the learning rate decays appropriately and each state-action pair  $\left( {s,a} \right)$  is executed an infinite number of times \cite{watkins1992q}. For real-world problems, the state space and/or the action space can be huge and it will take an excessive amount of time for  Q-learning to converge. 
	
	To allow fast convergence, the deep Q-network (DQN) algorithm uses a deep neural network model to approximate the Q-values in Q-learning \cite{DQNpaper}. For ease of exposition, we refer to the deep neural network model as DNN throughout this paper. The input to DNN is the agent's current state  $s$, and the outputs  are the approximated Q-values for different actions,  $\left\{ {Q\left( {s,a; \bm{\theta} } \right)|a \in A} \right\}$, where $\bm{\theta}$   is the neural network parameter and $A$  is the agent's action set. In time step  $t$, for action selection,  ${a_t} = \arg {\max _a}Q({s_t},a)$ in the  $\epsilon$-greedy algorithm is replaced by  ${a_t} = \arg {\max _a}Q({s_t},a;\bm{\theta} )$.
	
	For training of DNN, the parameter $\bm{\theta}$  is updated by minimizing the following loss function using a gradient descent method \cite{lecun2015deep}:\\
	\begin{equation}\label{eqn:loss1}
	\footnotesize
	L\left( \bm{\theta}  \right) = \frac{1}{{{N_E}}}\sum\limits_{{e_\tau } } {\left[ {{{\left( {{r_\tau } + \gamma  \mathop {\max }\limits_{a'} Q\left( {{s_{\tau  + 1}},a';{\bm{\theta'}}} \right) - Q\left( {{s_\tau },{a_\tau };\bm{\theta} } \right)} \right)}^2}} \right]}. 
	\end{equation}
	
	To stabilize the algorithm,  the ``experience replay'' \cite{lin1992self} and ``target neural network'' techniques are embedded into \eqref{eqn:loss1}. The details of these two techniques are given below.
	
	\underline{Experience Replay}: A first-in-first-out experience buffer is used to store a fixed number of experiences gathered from different time steps. Instead of training DNN with a single experience, multiple experiences are pooled together for batch training. Specifically, for a round of training, a minibatch of $N_E$  random experiences are sampled from the experience buffer in the computation of \eqref{eqn:loss1}, wherein the time index  $\tau$ denotes the time step at which that experience tuple ${e_\tau }$  was collected.  
	
	\underline{Target Neural Network}: A separate “target neural network” is used in the computation of ${r_\tau } + \gamma {\max _{a'}}Q\left( {{s_{\tau  + 1}},a';{{\bm{\theta'}}}} \right)$  in \eqref{eqn:loss1}. In particular, the target neural network's parameter vector is ${\bm{\theta'}}$  rather than  ${\bm{\theta}}$ in the DNN being trained. This separate target neural network is named target DNN and is a copy of DNN. The parameter ${\bm{\theta'}}$  of target DNN is updated to the latest  ${\bm{\theta} }$ of DNN every $F$  time steps, while the parameter ${\bm{\theta}}$  is updated every time step by updating the loss function \eqref{eqn:loss1}.  \\
	
	\section{System Model}\label{sec:system_model}
	\subsection{Heterogeneous Wireless Networks}
\begin{figure}[!t]
		\centering
		\includegraphics[scale=0.48]{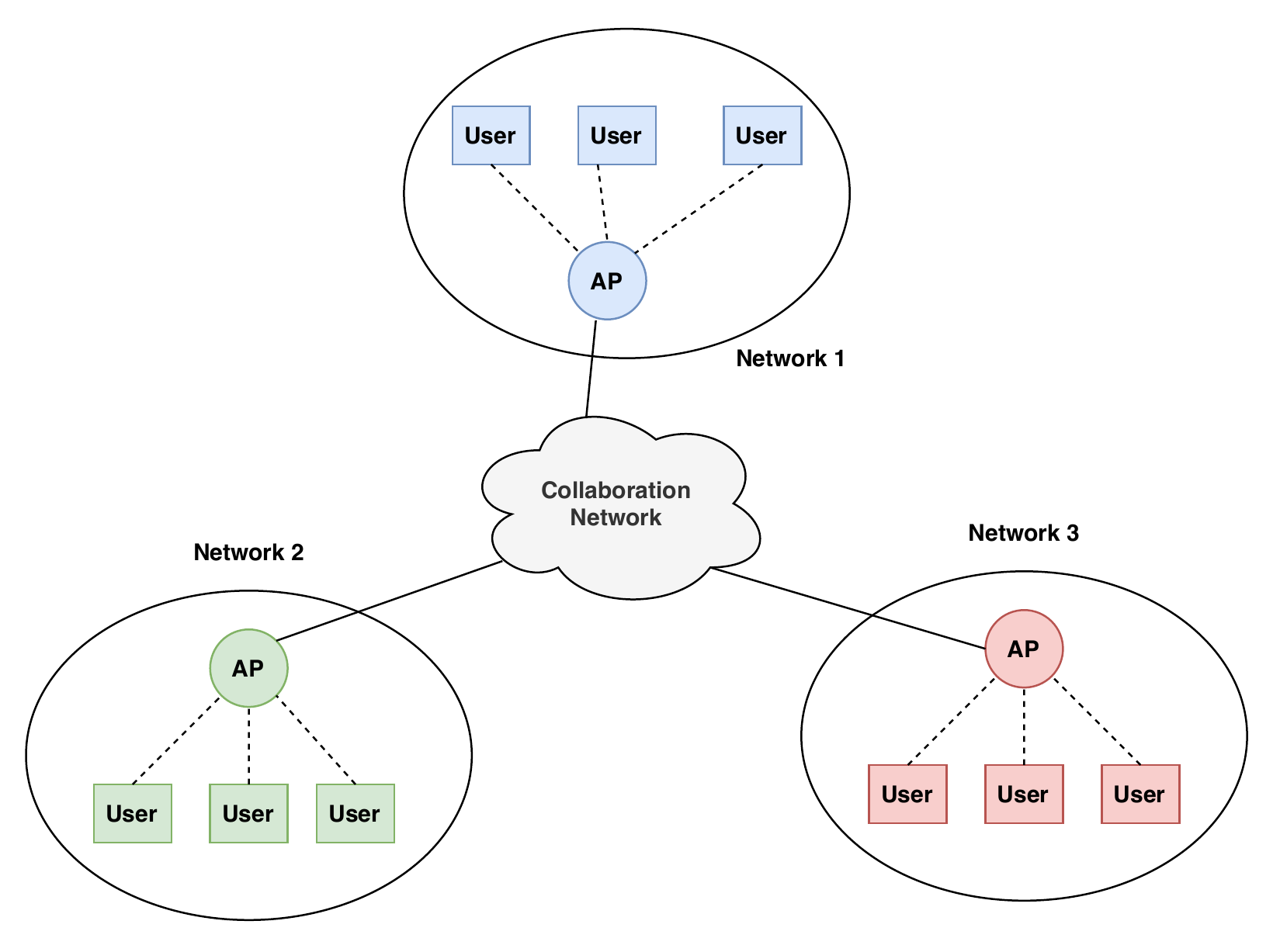}
		\caption[]{Heterogeneous wireless networks.}
		\label{fig:system_model}
	\end{figure}

	As illustrated in Fig. \ref{fig:system_model}, we  consider heterogeneous wireless networks, where multiple separate wireless networks share a common wireless spectrum to transmit data packets in a time-slotted manner.  These wireless networks are heterogeneous in that they may use different MAC protocols. Within each wireless network, multiple users adopt the same MAC protocol to transmit data packets to an access point (AP). The APs belonging to different networks are connected to a collaboration network \cite{tilghman2019will}. The collaboration network is a control network that is separate from the wireless networks, and allows different wireless networks to communicate collaborative information at a high level, e.g., the transmission results of different users in different wireless networks.  
	
	The focus of this paper is to design a MAC protocol for a particular wireless network. The underpinning technique of our MAC protocol is deep reinforcement learning. We refer to our MAC protocol as Deep-reinforcement Learning Multiple Access (DLMA) protocol. This particular wireless network is referred to as the DLMA network, and the users within the DLMA network are referred to as DLMA users. 
	
	The objective of the DLMA network is to achieve efficient and equitable spectrum allocation among different wireless networks. To this end, we adopt ``$\alpha$-fairness'' \cite{mo2000fair} as the general objective of the DLMA network. The DLMA network can adjust the value of parameter $\alpha$  to achieve different specific objectives. For example,  $\alpha  = 0$ corresponds to maximizing the sum throughput of all the wireless networks;  $\alpha  = 1$ corresponds to achieving proportional fairness;  $\alpha  \to \infty$ corresponds to achieving max-min fairness. The detailed formulation of the  $\alpha$-fairness objective is given in Section \ref{sec:fairness}. 
	
	\subsection{$\alpha$-Fairness Objective}\label{sec:fairness}
	We assume there are $L$  DLMA users in the DLMA network and $N-L$  non-DLMA users in other networks. We index the DLMA users by $i = 1,2, \ldots ,L$   and index the non-DLMA users by   $i = L + 1, \ldots ,N$. We define the ``\textbf{throughput}'' of a user as the average successful packet rate. Particularly, we use ${x^{\left( i \right)}}$  to denote the throughput of user  $i$, and if user $i$  successfully transmitted  $R$ packets within   $W$ time slots, then ${x^{\left( i \right)}}$  is calculated as  ${x^{\left( i \right)}} = R/W$.
	
	According to the  $\alpha$-fairness objective \cite{mo2000fair}, the utility function of user  $i$,  ${f_\alpha }\left( {{x^{\left( i \right)}}} \right)$, is given by
	\begin{align}
	{f_\alpha }\left( {{x^{\left( i \right)}}} \right) = \left\{ {\begin{array}{*{20}{c}}
		{{{\left( {{x^{\left( i \right)}}} \right)}^{1 - \alpha }}/\left( {1 - \alpha } \right),}&{\alpha  > 0{\rm{ \  \&  \ }}\alpha  \ne {\rm{1},}}\\
		{\log \left( {{x^{\left( i \right)}}} \right),}&{\alpha  = 1.}
		\end{array}} \right.
	\end{align}
	
	The overall objective of the DLMA network is to maximize the sum of the utility functions of all the users:
	\begin{align}
	\begin{array}{l}
	\text{maximize:} \ F\left( {{x^{\left( 1 \right)}}, \ldots ,{x^{\left( N \right)}}} \right) = \sum\limits_{i = 1}^N {{f_\alpha }\left( {{x^{\left( i \right)}}} \right)}  \\
	\text{subject to:} \ {x^{\left( i \right)}} \ge 0{\rm{ \  \&  \ }}\sum\limits_{i = 1}^N {{x^{\left( i \right)}}}  \le 1.
	\end{array}
	\label{eqn:fairness}
	\end{align}
	
	\subsection{AP-User Pair in the DLMA Network}
\begin{figure}[!t]
		\centering
		\includegraphics[scale=0.55]{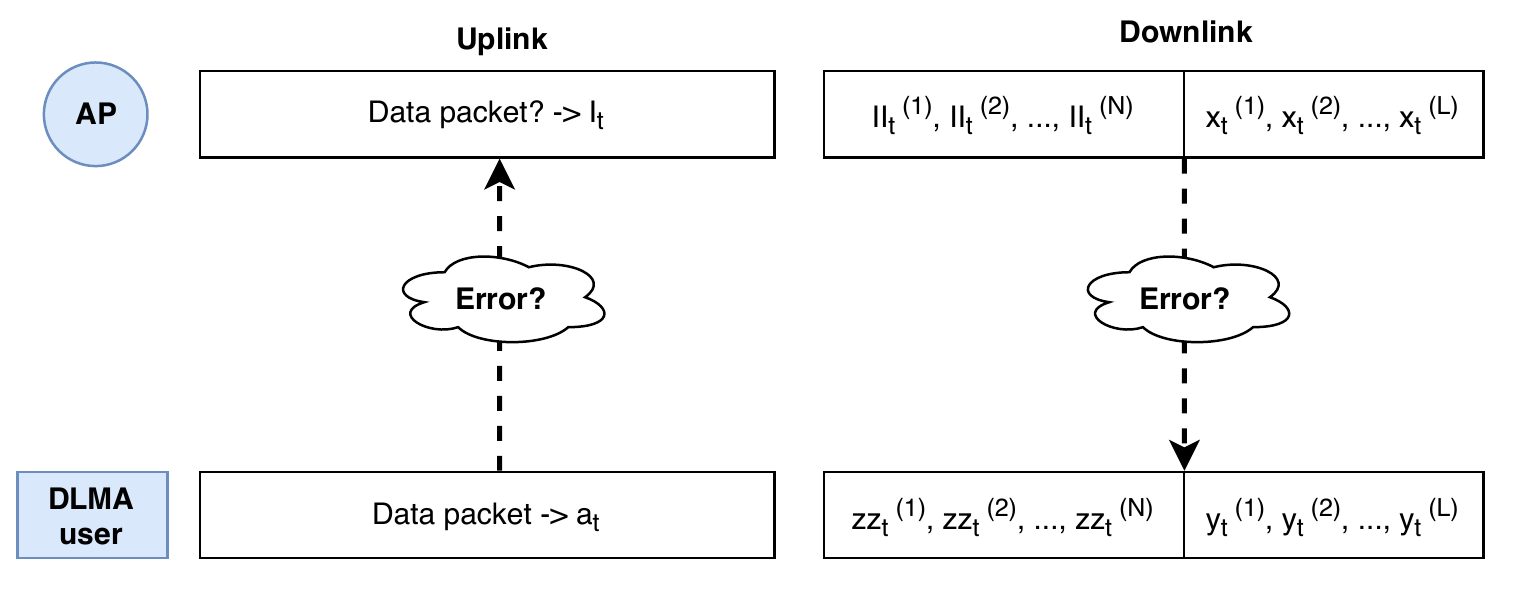}
		\caption[]{Uplink and downlink phases between the DLMA AP and one DLMA user.}
		\label{fig:up_down}
	\end{figure}

	We now describe the operations between one particular DLMA user and the AP in the DLMA network. Within each time slot, there are two phases between the DLMA user and the DLMA AP: the uplink phase for transmitting a data packet and the downlink phase for transmitting an ACK packet, as illustrated in Fig. \ref{fig:up_down}. We assume both the feedforward channel in the uplink phase and the feedback channel in the downlink phase are \textit{packet erasure channels} \cite{bertsekas1992data} with erasure probabilities  ${e_{up}}$ and  $e_{down}$, respectively. Specifically, the data packet from the DLMA user to the AP is lost with probability  ${e_{up}}$, and the ACK packet from the AP to the DLMA user is lost with probability  $e_{down}$. The details of these two phases are given below. 
	
	\subsubsection{Uplink Phase}
	In the uplink phase, at the beginning of each time slot, the MAC module in the DLMA user decides whether to transmit a data packet to the DLMA AP or not (the MAC decision making will be detailed in Section \ref{sec:protocol_design}). Specifically, ${a_t}$  is the decision variable for time slot  $t$, with ${a_t} = 1$  if the DLMA user transmits, and  ${a_t} = 0$  if the DLMA user does not transmit. At the AP side, the indicator variable  $I_t$ contains the outcome, with $I_t = S$ if a data packet from the DLMA user is received, and  $I_t = F$  otherwise. Note that three possibilities can lead to $I_t = F$: (i) the DLMA user transmitted a data packet, but the data packet is corrupted by channel noise; (ii) the DLMA user transmitted a data packet, but other users also transmitted in the same time slot, leading to a collision; (iii) the DLMA user did not transmit. 
	
	\subsubsection{Downlink Phase}
	In the downlink phase, the DLMA AP broadcasts an ACK packet (or feedback packet) to all the DLMA users. Even if no DLMA user transmits, the DLMA AP will still broadcast an ACK to indicate that it did not receive anything from the DLMA users in the time slot that has just transpired. As shown in Fig. \ref{fig:up_down}, the ACK packet includes two parts. The first part summarizes the transmission results of all the users. Specifically, at the end of time slot  $t$, the first part of the ACK packet contains   $\left( {II_t^{\left( 1 \right)}, \ldots ,II_t^{\left( L \right)}, \ldots ,II_t^{\left( N \right)}} \right)$,  where  $II_t^{\left( i \right)} = \left[ {I_t^{\left( i \right)},I_{t - 1}^{\left( i \right)}, \ldots ,I_{t - K + 1}^{\left( i \right)}} \right]$ ($i \in \left\{ {1,2, \cdots ,N} \right\}$) with $K$  being the history length of the transmission results  maintained by the AP. The reason for keeping $K$  transmission results in the ACK is to compensate for the possibility of lost ACKs, hence the loss of feedback, in earlier time slots  (see Section \ref{sec:neural_net_train} for details). The DLMA AP can construct  $II_t^{\left( i \right)}$ for $i \in \left\{ {1,2, \cdots ,L} \right\}$  based on the transmission results of the DLMA users and can obtain  $II_t^{\left( i \right)}$ for $i \in \left\{ {L + 1, \ldots ,N} \right\}$  from the collaboration network. At the DLMA user side, we use  $\left( {zz_t^{\left( 1 \right)}, \ldots ,zz_t^{\left( L \right)}, \ldots ,zz_t^{\left( N \right)}} \right)$ to represent the transmission results known to the DLMA user.  Specifically, if the ACK packet is successfully received, then $zz_t^{\left( i \right)} = II_t^{\left( i \right)}$ for all $i \in \left\{ {1,2, \cdots ,N} \right\}$; if the ACK packet is lost,  $zz_t^{\left( i \right)} = null$  for all  $i \in \left\{ {1,2, \cdots ,N} \right\}$. 
	
	The second part of the ACK packet is the throughput vector $\left[ {x_t^{\left( 1 \right)},x_t^{\left( 2 \right)}, \ldots ,x_t^{\left( L \right)}} \right]$ of all the DLMA users (here, $x_t^{\left( i \right)}$, $i \in \left\{ {1,2, \cdots ,L} \right\}$, is the throughput of DLMA user $i$ in time slot $t$). At the DLMA user side, we use $ \left[ {y_t^{\left( 1 \right)},y_t^{\left( 2 \right)}, \ldots ,y_t^{\left( L \right)}} \right] $ to represent the throughput  results known to the DLMA users. Specifically, if the ACK packet is successfully received, then  $y_t^{\left( i \right)} = x_t^{\left( i \right)}$  for any   $i \in \left\{ {1,2, \cdots ,L} \right\}$; if the ACK packet is lost, then  $y_t^{\left( i \right)} = y_{t - 1}^{\left( i \right)}$ for any  $i \in \left\{ {1,2, \cdots ,L} \right\}$. We will describe how our algorithm uses the throughput vector $\left[ {y_t^{\left( 1 \right)},y_t^{\left( 2 \right)}, \ldots ,y_t^{\left( L \right)}} \right]$ in Section \ref{sec:two_stage}. 
	
	\section{DLMA Protocol Design with DRL}\label{sec:protocol_design}
	This section describes the design of our DLMA protocol. We first provide the definitions for agent, action, observation, reward, and state. Then we present our proposed DLMA algorithm. 
	
	\subsection{Definitions of Agent, Action, Observation, Reward, and State}\label{sec:definitions}
	\underline{Agent}: We consider each DLMA user to be an agent in the parlance of machine learning. Each of the DLMA users makes its own transmission decision. We assume the DLMA users can only communicate with the DLMA AP and they do not have direct communications with each other.
	
	In our previous work \cite{yu2019deep} where the channels between the DLMA users and the DLMA AP were assumed to be perfect, the DLMA AP was regarded as a centralized agent responsible for making decisions for all DLMA users. In \cite{yu2019deep}, the DLMA AP broadcasts the control information containing the decisions to the DLMA user, telling them who should transmit in what slots. However, when the channels between the DLMA users and the DLMA AP are imperfect, the control information may be lost.  Without immediate and correct control information in a particular time slot, all users will not transmit even if the AP intends one user to transmit, resulting in significant performance degradation. 
	
	Unlike \cite{yu2019deep}, this paper adopts a distributed DLMA algorithm to solve the imperfect channel problem. In the distributed algorithm, all the DLMA users (i.e., the agents), make decisions on their own. Therefore, our problem is a multi-agent problem. Even if the feedback ACK is lost in a particular time slot, a DLMA user may still decide to transmit. To avoid collisions between the DLMA users, we propose a \textit{two-stage action selection mechanism}  in Section \ref{sec:two_stage}---more exactly, we can only reduce collisions but not avoid them altogether when the feedback channels are imperfect, as will be detailed later. We further propose a \textit{feedback recovery mechanism} in Section \ref{sec:neural_net_train} to reduce the detrimental effect of the imperfect feedback channels. 
	
	\underline{Action}: The possible actions of each agent are ``to transmit'' and ``not to transmit''. For each agent,  the decision/action variable ${a_t} = 1$  and ${a_t} = 0$  represent ``to transmit'' and ``not to transmit'' in time slot $t$, respectively. In the following, we will focus on one particular agent. For ease of exposition, we omit the index of this agent.
	
	\underline{Observation}: After the execution of an action in time slot  $t$, the agent has an observation  $o_t$. If  ${a_t} = 0$,  then there are two possible observations:   ${o_t} = B$ or  ${o_t} = I$, indicating that the channel was used by other users or the channel was idle in time slot $t$, respectively.  If  ${a_t} = 1$ and the ACK packet was successfully received, then there are two possible observations:  ${o_t} = {I_t} = S$  or ${o_t} = {I_t} = F$, indicating that the data packet was successfully received or not successfully received by the AP, respectively. If ${a_t} = 1$  and the ACK packet was not successfully received, then  ${o_t} = null$. Table \ref{table:observation} summarizes all the five possible observations.
	\begin{table}[htbp]
		\centering
		\caption{Possible observations of each agent in each time slot.}
		\label{table:observation}
		\renewcommand\arraystretch{1.4}
		\begin{tabular}{|l|l|lll}
			\cline{1-2}
			$o_t$ & \textbf{Description}  \\ \cline{1-2}
			\textit{B} & ${a_t} = 0$; the channel is being used by other user(s)  \\ \cline{1-2}
			\textit{I} & ${a_t} = 0$; the channel is not used by any user   \\ \cline{1-2}
			\textit{S} & ${a_t} = 1$; data packet is successful;  ACK packet is successful  \\ \cline{1-2}
			\textit{F} & ${a_t} = 1$; data packet is unsuccessful; ACK packet is successful
			\\ \cline{1-2}
			\textit{null} & ${a_t} = 1$; ACK packet is unsuccessful  \\ \cline{1-2}
		\end{tabular}
	\end{table}

	\underline{Reward}: We define a reward for each user to indicate whether its data packet was successfully received. To achieve a global  $\alpha$-fairness objective, we assume each agent knows the rewards of all the users. Specifically, at the end of time slot  $t$,  each agent can find out the rewards of all users from  $\left( {zz_t^{\left( 1 \right)}, \ldots ,zz_t^{\left( N \right)}} \right)$. Let the reward vector maintained by the agent be   ${{\bf{r}}_t} = \left[ {r_t^{\left( 1 \right)}, \ldots ,r_t^{\left( N \right)}} \right]$. The individual reward of user  $i$ ($i \in \left\{ {1, \ldots ,N} \right\}$),  $r_t^{\left( i \right)}$,  in ${{\bf{r}}_t}$  is decided as follows.  If  $zz_t^{\left( i \right)} = null$, i.e., the ACK packet was not received, then   $r_t^{\left( i \right)} = null$.  If    $zz_t^{\left( i \right)} \ne null$, i.e., the ACK packet was successfully received, and the first element in $zz_t^{\left( i \right)}$ is   $S$, i.e.,  $I_t^{\left( i \right)} = S$, then    $r_t^{\left( i \right)} = 1$. If  $zz_t^{\left( i \right)} \ne null$, and the first element in  $zz_t^{\left( i \right)}$ is  $F$, i.e.,  $I_t^{\left( i \right)} = F$,  then   $r_t^{\left( i \right)} = 0$. Overall, $r_t^{\left( i \right)}$   can be decided as follows:
	\begin{align}
	r_t^{\left( i \right)} = \left\{ {\begin{array}{*{20}{c}}
		{null,}&{\text{if} \ zz_t^{\left( i \right)} = null},\\
		{1,}&{\text{if} \ zz_t^{\left( i \right)} \ne null \ \& \ I_t^{\left( i \right)} = S},\\
		0&{\text{if} \ zz_t^{\left( i \right)} \ne null\ \& \ I_t^{\left( i \right)} = F}.
		\end{array}} \right.
	\end{align}
		
	We can see that the loss of the ACK packet results in $r_t^{\left( i \right)}$  being \textit{``null''} and there is no indication whether the data packet was successfully received or not. We say that $r_t^{\left( i \right)}$  is erroneous if the ACK is lost. Erroneous rewards will not ``advise''/``reinforce'' the agent correctly, and may lead to the wrong solution. We describe how we tackle this problem with the feedback recovery mechanism in Section \ref{sec:neural_net_train}. 

	\underline{State}: We define the ``channel state'' of the agent in time slot $t+1$  as  ${c_{t + 1}} = \left[ {{a_t},{o_t},r_t^{\left( 1 \right)}, \ldots ,r_t^{\left( N \right)}} \right]$. The channel state captures the action of the agent and the results of the action in the past time slot. We further define the ``state'' of the agent in time slot  $t+1$ as a concatenation of the past $M$  channel states, i.e.,  ${s_{t + 1}} = \left[ {{c_{t + 1}},{c_t}, \ldots ,{c_{t - M + 2}}} \right]$.
	
	Two things need to be pointed out here. The first is that the state of the agent may be ``noisy'' since the rewards of the agent can be erroneous due to imperfect feedback. The second is that due to  local actions and local observations, the states of different agents may be different in each time slot. The discrepancies in the states kept by different agents pose a challenge for these agents to come up with an overall coherent action plan. For example, the discrepancies may result in two agents deciding to transmit at the same time in the next time slot, resulting in a collision. This is the reason why we put forth a two-stage action selection mechanism to reduce collisions among the agents in Section \ref{sec:two_stage}.
	
\begin{figure}[!t]
		\centering
		\includegraphics[scale=0.9]{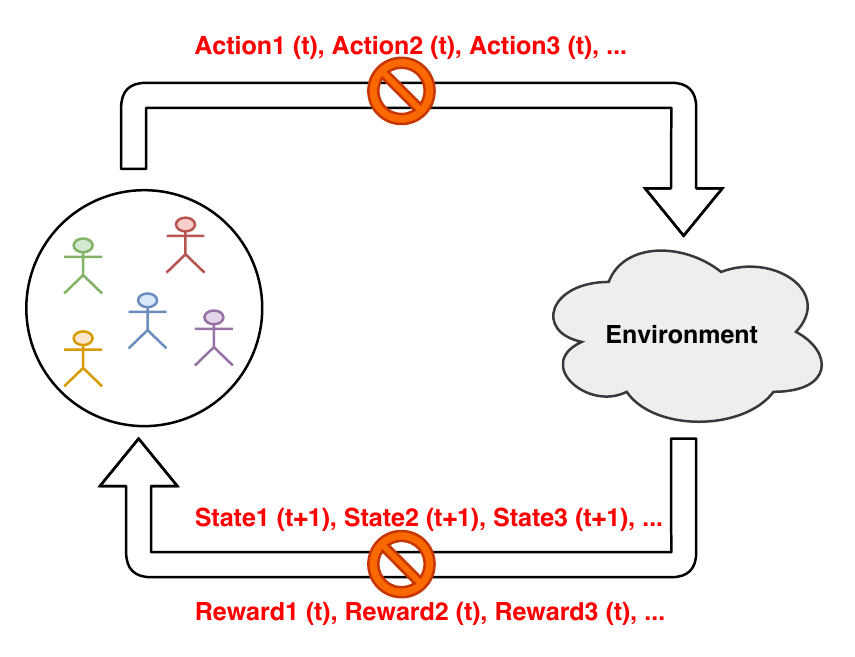}
		\caption[]{The modified reinforcement learning framework: multi-agent with imperfect channels.}
		\label{fig:RL_framework2}
	\end{figure}

	With the above definitions, we have formulated our multiple access problem into a multi-agent reinforcement learning problem. The modified reinforcement learning framework is illustrated in Fig. \ref{fig:RL_framework2}. In the following, we present our DLMA algorithm for solving the multi-agent problem.
	
	\subsection{DLMA Algorithm}
	The DLMA algorithm consists of two parts: (i) two-stage action selection; (ii) neural network training. An agent employs the two-stage action selection process to decide its action. Each agent keeps a deep neural network (DNN) model and trains it to optimize the collective policy of the agents from the agent's own perspective.  The details of these two parts are given below. 
	
	\subsubsection{Two-Stage Action Selection}\label{sec:two_stage}
	As illustrated in Fig. \ref{fig:two_stage}, in Stage 1 of the two-stage action selection process, an agent makes use of its DNN to come up with a decision for the overall DLMA network as to whether one of the agents should transmit. We refer to the decision for the overall DLMA network as ``\textbf{network action}''. If the network action in Stage 1 is to transmit, in Stage 2, the agent will then decide whether it is the one among all the agents that will perform the transmission. If the network action in Stage 1 is not to transmit, in Stage 2, the agent just does not transmit. We refer to the decision of the agent in Stage 2 as ``\textbf{agent action}''. We remark that the purpose of Stage 1 is to enable the DLMA network to achieve the  $\alpha$-fairness objective when coexisting with other networks, and the purpose of Stage 2 is to break ties among the agents in the DLMA network\footnote{Note that, as an alternative to the two-stage action selection process, each agent can also directly  generate its own action  based only on the outputs of its DNN, i.e., the DNN outputs the agent action directly rather than the network action, the implication of which is that the agent has to learn to coexist with the users in other networks and the other agents in the DLMA network simultaneously. Due to lack of coordination among the agents, it is more likely for  more than one agent to decide to transmit at the same time with this direct agent-action scheme. More frequent collisions among the DLMA agents may happen as a result. The above intuition has been borne out by our investigations. Specifically, we studied the coexistence of two agents with the direct-action generation approach and we found that collisions between these two agents will occur from time to time even though they have the same objective.}. The details of these two stages are given as follows:

	\begin{figure}[t]
		\centering
		\includegraphics[scale=0.75]{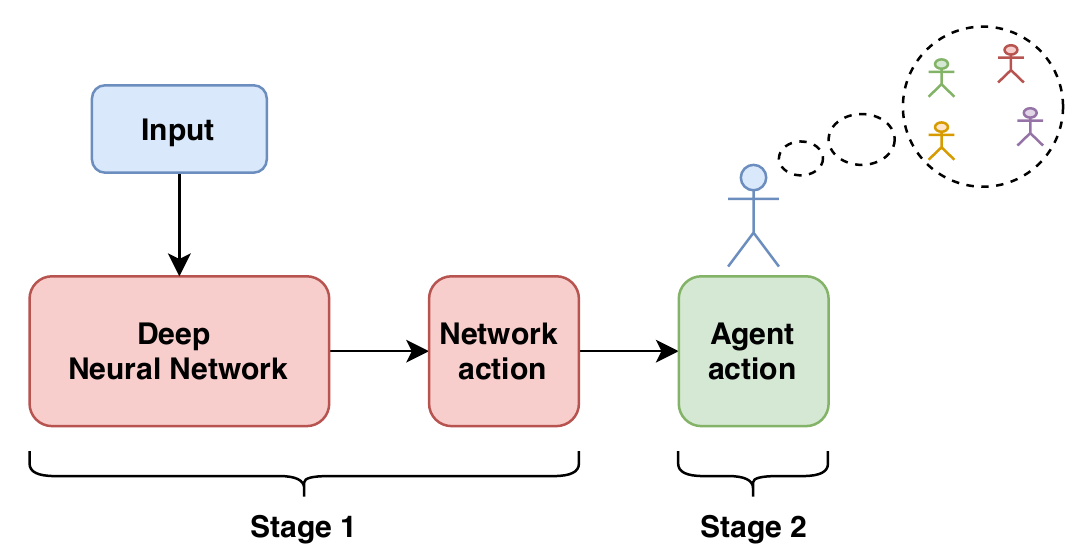}
		\caption[]{An illustration of the two-stage action selection process.}
		\label{fig:two_stage}
	\end{figure}
	
	\underline{Stage 1}: At the beginning of  time slot  $t$, the agent inputs its current state $s_t$  into the DNN, and the DNN outputs multiple Q values  ${\bf{Q}} = \left\{ {{Q^{\left( i \right)}}\left( {{s_t},u;\bm{\theta} } \right)|i \in \left\{ {0,L + 1, \ldots ,N} \right\},u \in \left\{ {0,1} \right\}} \right\}$, where $u$  is the network action that needs to be decided, ${Q^{\left( 0 \right)}}\left( {{s_t},u;\bm{\theta} } \right) \approx {\bf{E}}\left[ {\sum\limits_{k = 0}^\infty  {\sum\limits_{i = 1}^L {{\gamma ^k} \cdot r_{t + k}^{\left( i \right)}} } } \right]$  is the estimated cumulative discounted rewards of the DLMA network, and  ${Q^{\left( i \right)}}\left( {{s_t},u;\bm{\theta} } \right) \approx {\bf{E}}\left[ {\sum\limits_{k = 0}^\infty  {{\gamma ^k} \cdot r_{t + k}^{\left( i \right)}} } \right]$ ($i \in \left\{ {L + 1, \ldots ,N} \right\}$) is the estimated cumulative discounted rewards of the non-DLMA user  $i$. Based on the Q values, the agent can decide the network action $u_t$ according to \eqref{eqn:greedy} on page 9.
	
	In \eqref{eqn:greedy},  ${Q^{\left( 0 \right)}}\left( {{s_t},u;\bm{\theta} } \right)/L$ is the estimated cumulative discounted rewards of each agent, and  $L \cdot {f_\alpha }\left( {{Q^{\left( 0 \right)}}\left( {{s_t},u;\bm{\theta} } \right)/L} \right)$ can be regarded as the sum utilities of all the agents (i.e., the utility of the DLMA network). Meanwhile, $\sum\nolimits_{i = L + 1}^N {{f_\alpha }\left( {{Q^{\left( i \right)}}\left( {{s_t},u;\bm{\theta} } \right)} \right)}$ can be regarded as the sum utilities of all the non-agent users.

	\underline{Stage 2}: In time slot  $t$, the agent action $a_t$  of agent  $l$ ($l \in \left\{ {1, \ldots ,L} \right\}$) is determined by the network action  $u_t$ and the throughput vector $\left[ {y_{t - 1}^{\left( 1 \right)},y_{t - 1}^{\left( 2 \right)}, \ldots ,y_{t - 1}^{\left( L \right)}} \right]$  of all the agents. Specifically, if  ${u_t} = 1$, i.e., the network action is to transmit, and the throughput vector satisfies \eqref{eqn:throughput_vector} on page 9,  then  ${a_t} = 1$, i.e., the agent action is to transmit; otherwise,   ${a_t} = 0$, i.e., the agent action is not to transmit.
	
	The interpretation of  \eqref{eqn:throughput_vector}  is as follows: the throughput of agent  $l$ is the smallest one among the throughputs of all the agents, and the throughputs of those agents with indexes  $i < l$ are larger than the throughput of the agent  $l$. \\

	\subsubsection{Neural Network Training}\label{sec:neural_net_train}
	We next describe the training of the DNN used in Stage 1 of the two-stage action selection process. In the original DQN algorithm \cite{DQNpaper} where the feedback from the environment is assumed to be perfect and without errors, the agent stores the experiences in the form of \textit{(state, action, reward, next state)} in an experience buffer, and samples multiple experiences in each time step to calculate a loss function and update the DNN using a gradient descent method (see the introduction of DQN in Section \ref{sec:DQN}). 
	
	As in \cite{DQNpaper}, here we also define an experience in the form of \textit{(state, action, reward, next state)} in our DLMA algorithm. A key difference is that we use a reward vector rather than a scalar reward as in \cite{DQNpaper}. Particularly, the experience collected in time slot  $t$ is represented by  ${e_t} = \left( {{s_t},{u_t},r_t^{\left( 1 \right)}, \ldots ,r_t^{\left( N \right)},{s_{t + 1}}} \right)$. Note that in the experience $e_t$ , we use the network action $u_t$  rather than the agent action  $a_t$ since we use the  DNN to generate the network action rather than the agent action.
	
	\begin{figure}[!t]
		\centering
		\includegraphics[scale=0.6]{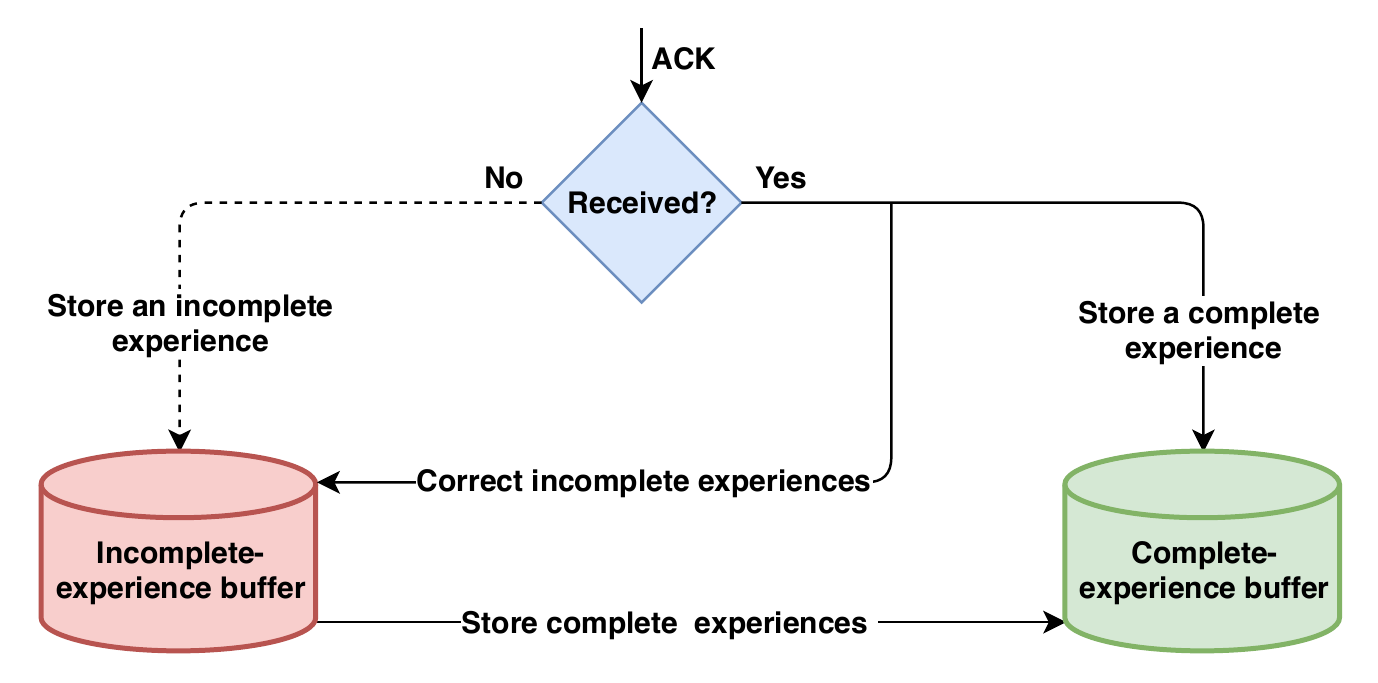}
		\caption[]{Feedback recovery mechanism: experience replay with two experience buffers.}
		\label{fig:experience_replay}
	\end{figure}

	Unlike the problem in \cite{DQNpaper}, in our problem, the feedback information from the environment may be erroneous and incomplete due to imperfect feedback channels. Particularly, when the ACK packet is lost in time slot  $t$, as explained in Section \ref{sec:definitions}, the reward $r_t^{\left( i \right)}$  for all  $i \in \left\{ {1, \ldots ,N} \right\}$ in  $e_t$ is \textit{``null''}. The experience  $e_t$ is deemed to be incomplete in this case. When the ACK packet is successfully received, $r_t^{\left( i \right)}$ is not \textit{``null''}, and the experience $e_t$  is deemed to be complete. Incomplete experiences will not ``advise''/``reinforce'' the agents correctly. Therefore, incomplete experiences will not be used to train the DNN. However, it is possible to fill in the missing rewards in the incomplete experiences via ACKs received later. Once the missing reward values are filled with the correct reward values, an incomplete experience will become an complete experience that can be used to train the DNN. We refer to this process of making incomplete experience complete as a  \textbf{feedback recovery mechanism}. 
	
	In the feedback recovery mechanism, instead of using one experience buffer, we use two experience buffers for each agent: an incomplete-experience buffer for storing the incomplete experiences and a complete-experience buffer for storing the complete experiences. The incomplete experiences will not be used to train the DNN. Only the complete experiences are sampled to calculate the loss function (will be given later) and train the DNN. If the information about the rewards of an incomplete experience is obtained through an ACK packet that arrives later (recall that the ACK packet contains transmission results up to  $K$ time slots), then the experience can become complete and will be put into the complete-experience buffer. Fig. \ref{fig:experience_replay} illustrates the operation of the feedback recovery mechanism and an example is given below to explain the details.
	
	For each agent, if the ACK packet is lost at the end of time slot  $t$, then the experience   ${e_t} = \left( {{s_t},{u_t},r_t^{\left( 1 \right)}, \ldots ,r_t^{\left( N \right)},{s_{t + 1}}} \right)$ is stored into the incomplete-experience buffer. In this case,  $r_t^{\left( i \right)} = null$ for all  $i \in \left\{ {1,2, \ldots ,N} \right\}$. 
	
	If the ACK packet is successfully received, then for all  ${e_{t - \Delta t}} = \left( {{s_{t - \Delta t}},{u_{t - \Delta t}},r_{t - \Delta t}^{\left( 1 \right)}, \ldots ,r_{t - \Delta t}^{\left( N \right)},{s_{t - \Delta t + 1}}} \right)$,  $1 \le \Delta t \le K - 1$, in the incomplete-experience buffer (if there is no experience in the incomplete-experience buffer, then do nothing), replace  $r_{t - \Delta t}^{\left( i \right)}$  for all $i \in \left\{ {1,2, \ldots ,N} \right\}$ with $\hat r_{t - \Delta t}^{\left( i \right)}$  and move ${\hat e_{t - \Delta t}} = \left( {{s_{t - \Delta t}},{u_{t - \Delta t}},\hat r_{t - t}^{\left( 1 \right)}, \ldots ,\hat r_{t - t}^{\left( N \right)},{s_{t - \Delta t + 1}}} \right)$  to the complete-experience buffer. In particular,  $\hat r_{t - \Delta t}^{\left( i \right)}$ is decided as follows: if  $I_{t - \Delta t}^{\left( i \right)} = S$ in  $zz_t^{\left( i \right)} = II_t^{\left( i \right)} = \left[ {I_t^{\left( i \right)}, \ldots ,I_{t - \Delta t}^{\left( i \right)}, \ldots ,I_{t - K + 1}^{\left( i \right)}} \right]$, then $\hat r_{t - \Delta t}^{\left( i \right)} = 1$; if  $I_{t - \Delta t}^{\left( i \right)} = F$, then  $\hat r_{t - \Delta t}^{\left( i \right)} = 0$. After that,  we clear the incomplete-experience buffer. For the newly collected experience  ${e_t} = \left( {{s_t},{u_t},r_t^{\left( 1 \right)}, \ldots ,r_t^{\left( N \right)},{s_{t + 1}}} \right)$, for consistency, we also add a hat on  $r_t^{\left( i \right)}$ for all  $i \in \left\{ {1,2, \ldots ,N} \right\}$, i.e.,  $\hat r_t^{\left( i \right)} = r_t^{\left( i \right)}$.  Then we store ${\hat e_t} = \left( {{s_t},{u_t},\hat r_t^{\left( 1 \right)}, \ldots ,\hat r_t^{\left( N \right)},{s_{t + 1}}} \right)$  into the complete-experience buffer.

	Note that when the ACK packet is successfully received at the end of time slot  $t$, the incomplete-experience buffer may still contain some incomplete experiences ${e_{t - \Delta t}} = \left( {{s_{t - \Delta t}},{u_{t - \Delta t}},r_{t - \Delta t}^{\left( 1 \right)}, \ldots ,r_{t - \Delta t}^{\left( N \right)},{s_{t - \Delta t + 1}}} \right)$  with $\Delta t \ge K$. In this case, the ACK packet with a window size of  $K$ time slots cannot correct the null rewards in these experiences. We discard these experiences by clearing the incomplete-experience buffer. The probability of  an experience being incomplete without being made complete is   ${e^K}$. When $K$ is large,  ${e^K} \to 0$.

	For training of DNN, multiple experiences are sampled from the complete-experience buffer to compute the loss function \eqref{eqn:loss2} on page 9, wherein ${u_{\tau  + 1}}$ is given by \eqref{eqn:next_action}. After computing the loss function \eqref{eqn:loss2}, the parameter $\bm{\theta}$  can be updated using a gradient descent method \cite{lecun2015deep}. Overall, the pseudocode of our DLMA algorithm is summarized in Algorithm \ref{alg:algorithm}. 

	\section{Performance Evaluation}
	The section evaluates the performance of our DLMA protocol. First, we present the simulation setup. After that, we examine the performance of DLMA in different scenarios: single agent with imperfect channels, multiple agents with perfect channels, and multiple agents with imperfect channels. 
	
	\subsection{Simulation Setup}
\begin{figure}[!t]
		\centering
		\includegraphics[scale=0.7]{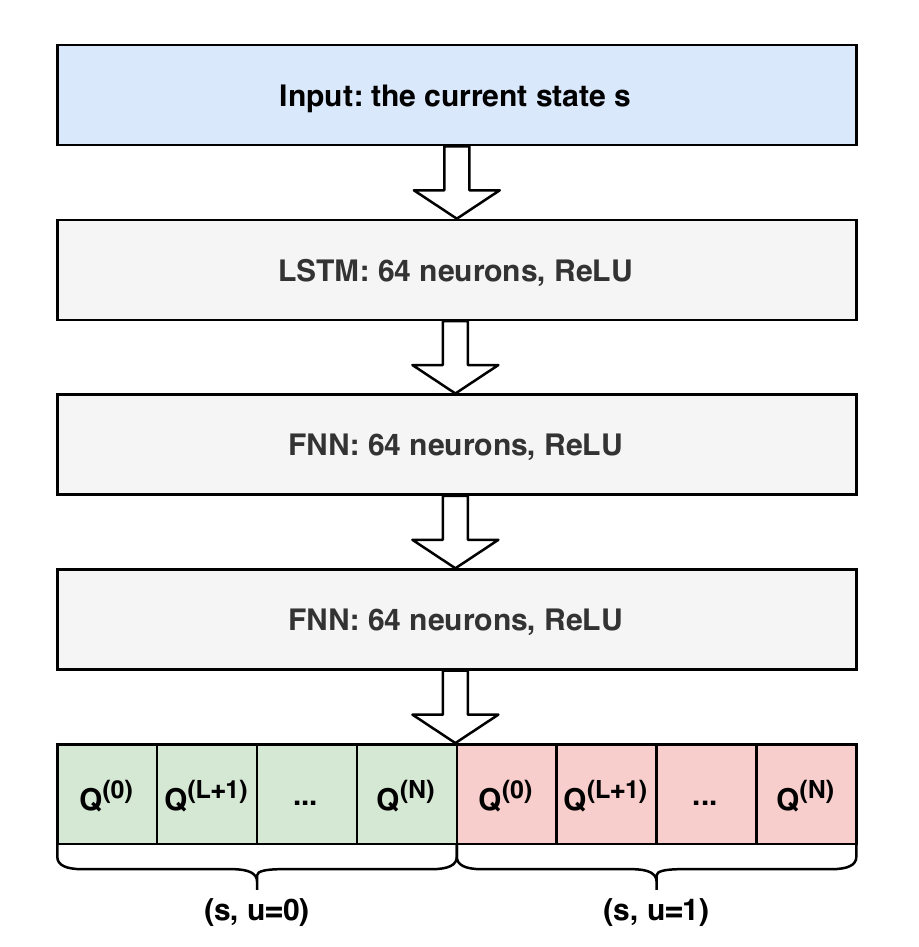}
		\caption[]{An illustration of the architecture of deep neural network model.}
		\label{fig:DNN}
	\end{figure}

	In our distributed DLMA algorithm, all the agents have the same objective and run the same algorithm simultaneously. Table \ref{table:hyperparameter} summarizes the hyperparameters adopted by each agent. In particular, the state of the agent contains  $M = 20$ channel states (see the definitions in Section \ref{sec:definitions}). The parameter  $\epsilon$ in the  $\epsilon$-greedy algorithm is initially set to 1 and is updated by  $\epsilon  \leftarrow \max \left\{ {0.995*\epsilon ,0.05} \right\}$ in each time slot. The size of the complete-experience buffer is 1000 and 64 experiences are sampled from the complete-experience buffer to perform gradient descent over the loss function \eqref{eqn:loss2} using the RMSProp algorithm \cite{lecun2015deep} in each time slot. The target network is updated once every 20 time slots. For the DNN, the input is the current state of the agent; the hidden layers consist of one long-short-term-memory (LSTM) \cite{hochreiter1997long} layer and two feedforward layers; the number of neurons in each hidden layer is 64 and the activation function is ReLU \cite{lecun2015deep}; the number of neurons in the output layer is   $2\left( {N - L + 1} \right)$. An illustration of the architecture of the DNN is presented in Fig. \ref{fig:DNN}. 
	
	\begin{table}[!t]
		\centering
		\caption{DLMA Hyperparameters.}
		\label{table:hyperparameter}
		\renewcommand\arraystretch{1.4}
		\begin{tabular}{|l|l|}
			\cline{1-2}
			\textbf{Hyperparameter} & \textbf{Value}  \\ \cline{1-2}
			State length   $M$ & 20  \\ \cline{1-2}
			$\epsilon$ in $\epsilon$-greedy algorithm & 1 to 0.05   \\ \cline{1-2}
			Discount factor $\gamma$ & 0.9 \\ \cline{1-2}
			Experience buffer size & 1000  \\ \cline{1-2}
			Experience-replay minibatch size & 64 \\ \cline{1-2}
			Target network update frequency & 1/20 \\ \cline{1-2}
		\end{tabular}
	\end{table}
	
	\subsection{Evaluation of DLMA in Different Scenarios}
	\subsubsection{Single Agent with Imperfect Channels}\label{sec:eva1}
	We now examine the ability of DLMA in dealing with the imperfect channel problem. In particular, we consider the coexistence of one DLMA agent with one TDMA user and one ALOHA user. The TDMA user is from the TDMA network and transmits in the \nth{2} slot out of 5 slots within a TDMA frame. The ALOHA user is from the ALOHA network and it transmits with a probability of 0.2 in each time slot.

	\begin{figure*}[t]
		\normalsize
		\begin{IEEEeqnarray}{rCl}
			{u_t} = \left\{ {\begin{array}{*{20}{c}}
					{\text{random selection in} \ \left\{ {0,1} \right\},}&{\ \text{with prob.} \ \epsilon,}\\
					{\mathop {\arg \max }\limits_{u \in \left\{ {0,1} \right\}} \left\{ {L \cdot {f_\alpha }\left( {\frac{{{Q^{\left( 0 \right)}}\left( {{s_t},u;\bm{\theta} } \right)}}{L}} \right) + \sum\limits_{i = L + 1}^N {{f_\alpha }\left( {{Q^{\left( i \right)}}\left( {{s_t},u;\bm{\theta} } \right)} \right)} } \right\}}&{\ \text{with prob.} \ 1 - \epsilon.} 
			\end{array}} \right.
			\label{eqn:greedy}
		\end{IEEEeqnarray}
		\vspace*{-0.1in}
	\end{figure*}
	\begin{figure*}[t]
		\normalsize
		\begin{IEEEeqnarray}{rCl}
			y_{t - 1}^{\left( l \right)} = \min \left\{ {y_{t - 1}^{\left( 1 \right)}, \ldots ,y_{t - 1}^{\left( L \right)}} \right\} \ \& \  y_{t - 1}^{\left( i \right)} > y_{t - 1}^{\left( l \right)} \  \text{for any} \  i < l.
			\label{eqn:throughput_vector}
		\end{IEEEeqnarray}
		\vspace*{-0.1in}
	\end{figure*}
	
	\begin{figure*}[t]
		\footnotesize
		\begin{IEEEeqnarray}{rCl}
			L\left( \bm{\theta}  \right) = \sum\limits_{{{\hat e}_\tau }} {\left\{ {{{\left( {\sum\limits_{i = 1}^L {\hat r_\tau ^{\left( i \right)}}  + \gamma  {Q^{\left( 0 \right)}}\left( {{s_{\tau  + 1}},{u_{\tau  + 1}};{\bm{\theta'}}} \right) - {Q^{\left( 0 \right)}}\left( {{s_\tau },{u_\tau };\bm{\theta} } \right)} \right)}^2} + \sum\limits_{i = L + 1}^N {{{\left( {\hat r_\tau ^{\left( i \right)} + \gamma {Q^{\left( i \right)}}\left( {{s_{\tau  + 1}},{u_{\tau  + 1}};{\bm{\theta'}}} \right) - {Q^{\left( i \right)}}\left( {{s_\tau },{u_\tau };\bm{\theta} } \right)} \right)}^2}} } \right\}}. 
			\label{eqn:loss2}
		\end{IEEEeqnarray}
		\vspace*{-0.1in}
	\end{figure*}
	\begin{figure*}[t]
		\normalsize
		\begin{IEEEeqnarray}{rCl}
			{u_{\tau  + 1}} = \mathop {\arg \max }\limits_{u' \in \left\{ {0,1} \right\}} \left\{ {L \cdot {f_\alpha }\left( {\frac{{{Q^{\left( 0 \right)}}\left( {{s_{\tau  + 1}},u';{\bm{\theta'}}} \right)}}{L}} \right) + \sum\limits_{i = L + 1}^N {{f_\alpha }\left( {{Q^{\left( i \right)}}\left( {{s_{\tau  + 1}},u';{\bm{\theta'}}} \right)} \right)} } \right\}. \label{eqn:next_action}
		\end{IEEEeqnarray}
	\end{figure*}
	
	\begin{algorithm*}[!t]
		\caption{Distributed DLMA algorithm}\label{alg:algorithm}
		\begin{algorithmic}
			\State \# All agents run the same algorithm simultaneously 
			\State Initialize state of the agent, the parameters of DNN and target DNN, and the hyperparameters
			\State Initialize the incomplete-experience buffer and the complete-experience buffer
			\For{$ t=1,2,\cdots $ in DLMA}
			\State Decide the network action $u_t$  and the agent action  $a_t$ according to the two-stage action selection process
			\State Interact with the environment with  $a_t$, and observe $zz_t^{\left( 1 \right)}, \ldots ,zz_t^{\left( N \right)}$,  $y_t^{\left( 1 \right)},y_t^{\left( 2 \right)}, \ldots ,y_t^{\left( L \right)}$
			\State Get to know  $r_t^{\left( 1 \right)}, \ldots ,r_t^{\left( N \right)}$  and  $o_t$ based on  $zz_t^{\left( 1 \right)}, \ldots ,zz_t^{\left( N \right)}$
			\State Construct  ${s_{t + 1}}$  based on  ${s_t},{a_t},{o_t},r_t^{\left( 1 \right)}, \ldots ,r_t^{\left( N \right)}$
			\If{$zz_t^{\left( i \right)} = null$ for $i \in \left\{ {1, \ldots ,N} \right\}$: \# ACK is not received}
			\State Store   ${e_t} = \left( {{s_t},{u_t},r_t^{\left( 1 \right)}, \ldots ,r_t^{\left( N \right)},{s_{t + 1}}} \right)$ into the incomplete-experience buffer
			\Else \ \# ACK is received
			\For{${e_{t - \Delta t}} = \left( {{s_{t - \Delta t}},{u_{t - \Delta t}},r_{t - \Delta t}^{\left( 1 \right)}, \ldots ,r_{t - \Delta t}^{\left( N \right)},{s_{t - \Delta t + 1}}} \right)$, $1 \le \Delta t \le K - 1$, in the incomplete-experience buffer}
			\State replace $r_{t - \Delta t}^{\left( i \right)}$  with $\hat r_{t - \Delta t}^{\left( i \right)}$  for  $i \in \left\{ {1, \ldots ,N} \right\}$
			\State Store ${\hat e_{t - \Delta t}} = \left( {{s_{t - \Delta t}},{u_{t - \Delta t}},\hat r_{t - \Delta t}^{\left( 1 \right)}, \ldots ,\hat r_{t - \Delta t}^{\left( N \right)},{s_{t - \Delta t + 1}}} \right)$  into the complete-experience buffer
			\EndFor 
			\State Clear the incomplete-experience buffer
			\State Set $\hat r_t^{\left( i \right)} = r_t^{\left( i \right)}$  for  $i \in \left\{ {1, \ldots ,N} \right\}$ and store ${\hat e_t} = \left( {{s_t},{u_t},\hat r_t^{\left( 1 \right)}, \ldots ,\hat r_t^{\left( N \right)},{s_{t + 1}}} \right)$  into the complete-experience buffer
			\EndIf
			
			\State \Call{\textbf{\textit{TrainDNN()}}}{}
			\State Every  $F$ time slots, update the parameters of the target DNN
			\EndFor \\
			\Procedure{\textbf{\textit{TrainDNN()}}}{}
			\State Sample multiple experiences from the complete-experience buffer
			\State Compute the loss function \eqref{eqn:loss2}
			\State Update the parameters of  DNN
			\EndProcedure
		\end{algorithmic}
	\end{algorithm*}
	We first study the performance of DLMA in reducing the detrimental effect of the imperfect feedback channels when the objective of the agent is to maximize sum throughput. For maximizing sum throughput, we set $\alpha = 0$ in the  $\alpha$-fairness objective. To study channels of varying imperfection, we set  the uplink channel erasure probability  $e_{up}$ to 0 and vary the downlink erasure probability   $e_{down}$. 
	
	For benchmarking, we replace the DLMA agent with a model-aware user. The model-aware user has the same objective as the agent and knows the transmission pattern of TDMA and the transmission probability of ALOHA. In addition, we assume the feedback channel of the model-aware user is perfect. A separate document \cite{benchmark} derives the benchmark for this scenario and the benchmarks for the scenarios in the subsequent sections. To conserve space, we omit the details of the derivations in this paper. 

	\begin{figure*}[!t]
		\centering
		\includegraphics[scale=0.75]{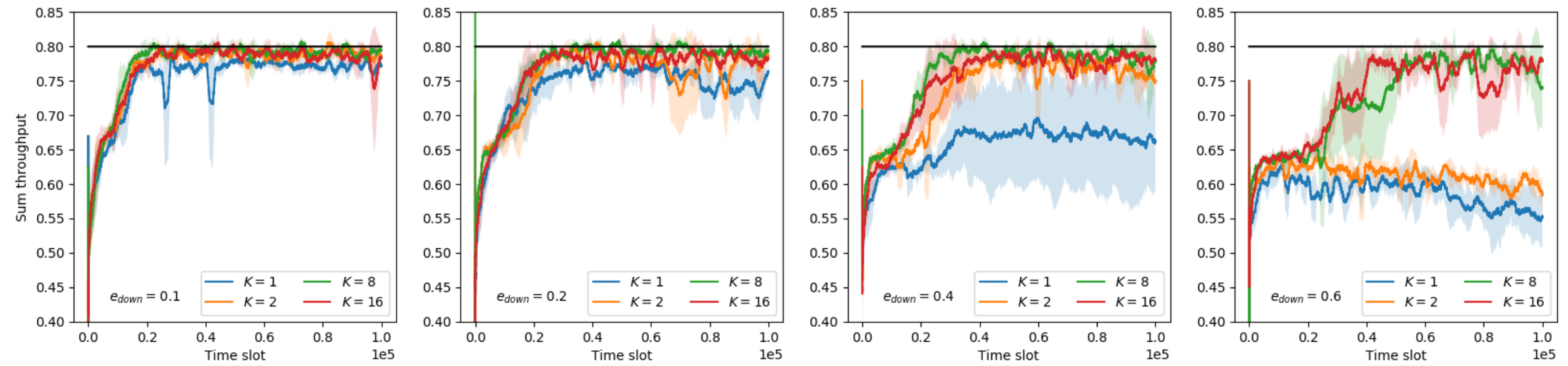}
		\caption[]{Sum throughput when one DLMA agent coexists with one TDMA user and one ALOHA user. The objective of the agent is to maximize sum throughput. From left to right, the downlink channel error probabilities are $e_{down} = 0.1, 0.2, 0.4, 0.6$.}
		\label{fig:eva1}
	\end{figure*}
	\begin{figure*}[!t]
		\centering
		\includegraphics[scale=0.75]{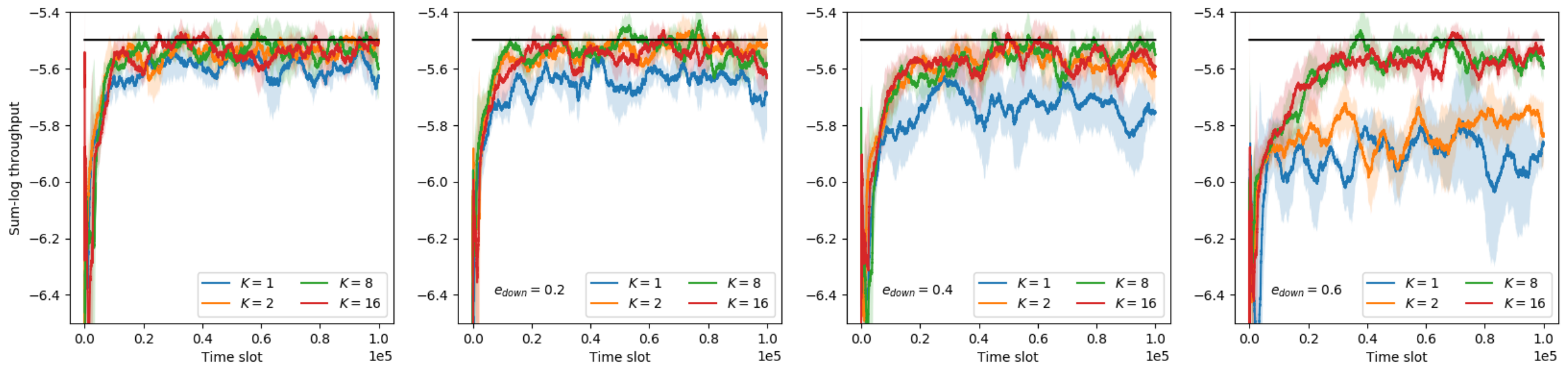}
		\caption[]{Sum-log throughput when one DLMA agent coexists with one TDMA user and one ALOHA user. The objective of the agent is to achieve proportional fairness. From left to right, the downlink channel error probabilities are $e_{down} = 0.1, 0.2, 0.4, 0.6$.}
		\label{fig:eva2}
	\end{figure*}

	\begin{figure}[!t]
		\centering
		\includegraphics[scale=0.43]{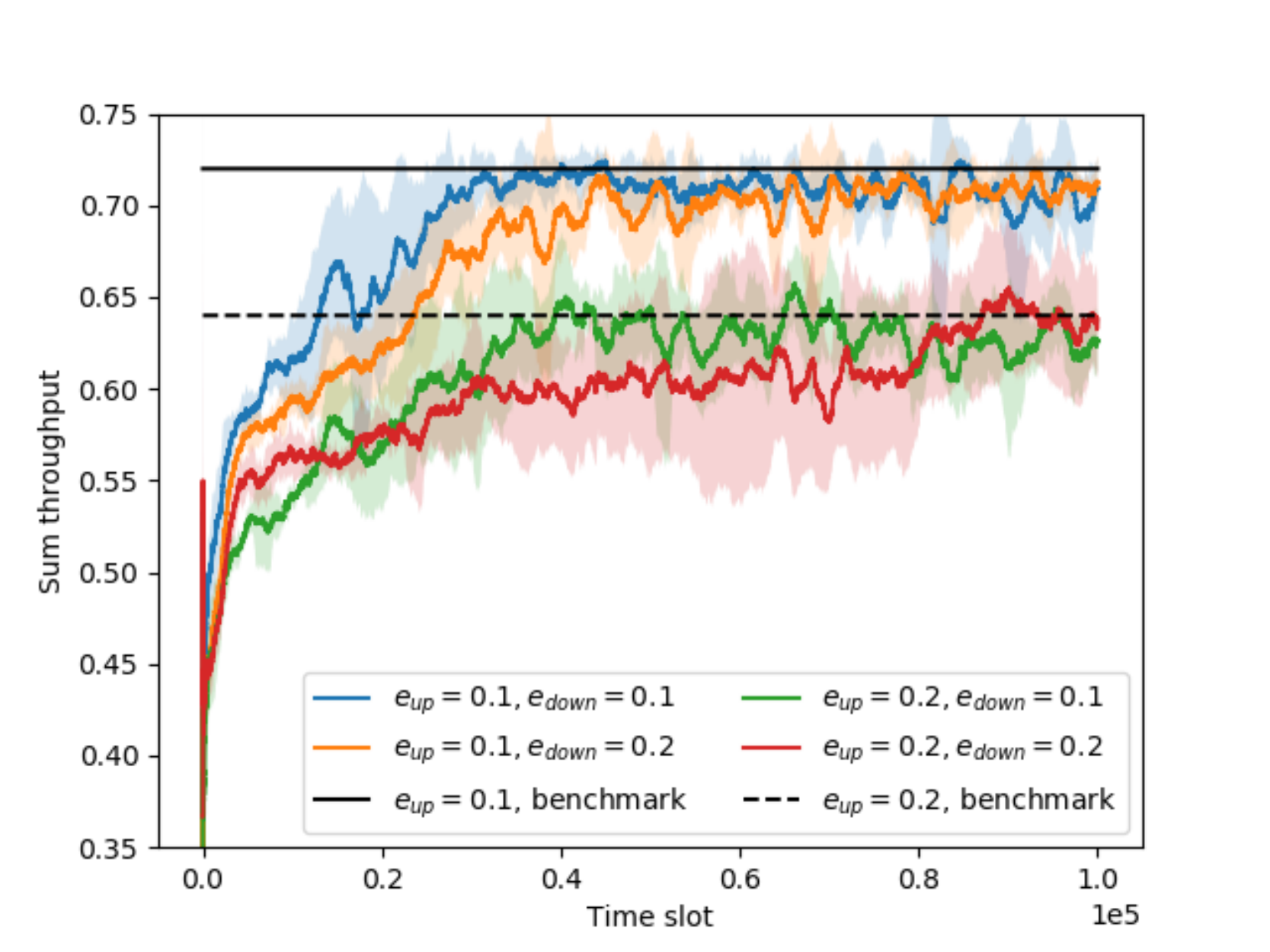}
		\caption[]{Sum throughput when one DLMA agent coexists with one TDMA user and one ALOHA user. The objective of the agent is to maximize sum throughput.}
		\label{fig:eva1_1}
	\end{figure}
	
	\begin{figure}[!t]
		\centering
		\includegraphics[scale=0.43]{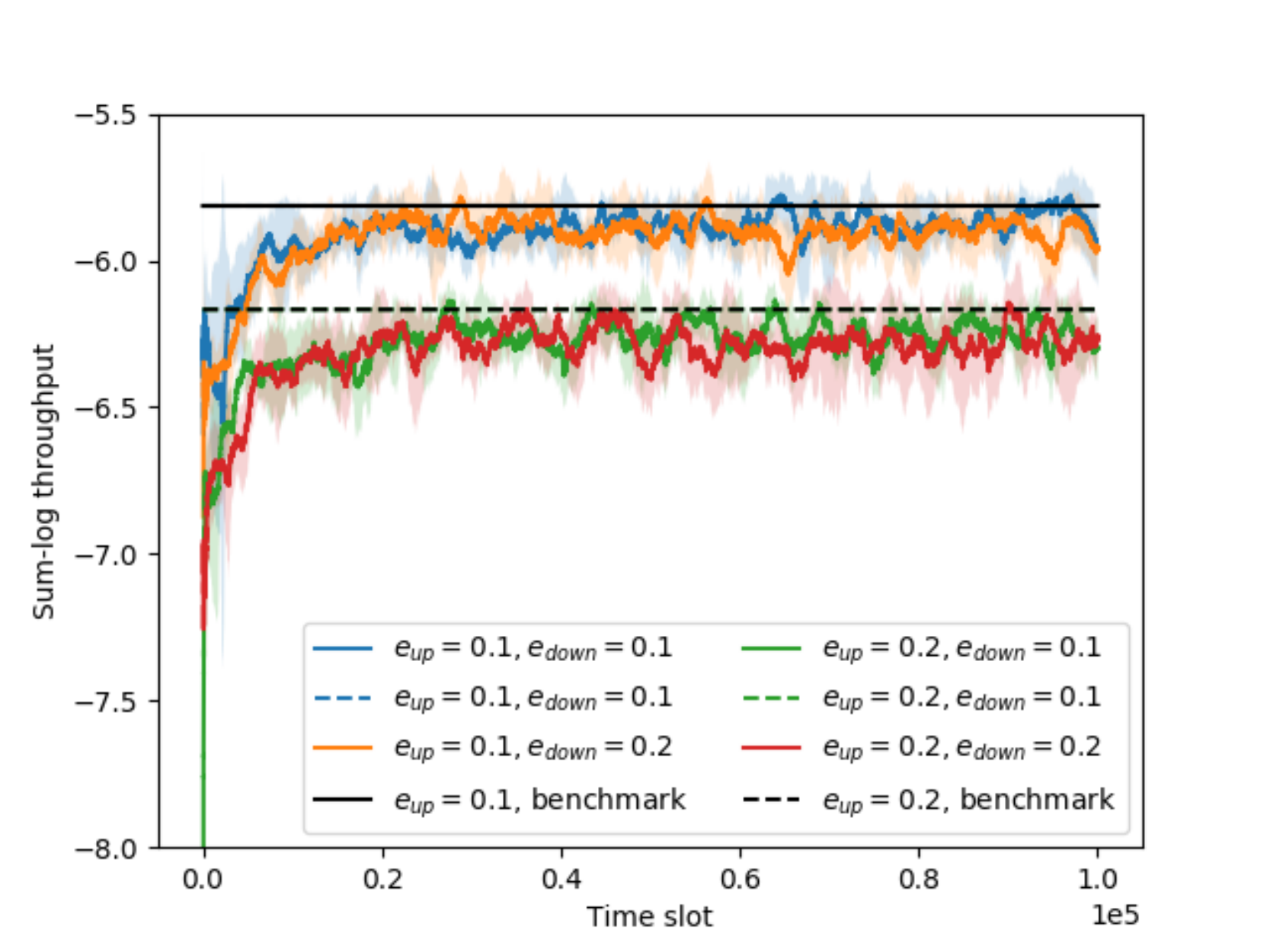}
		\caption[]{Sum-log throughput when one DLMA agent coexists with one TDMA user and one ALOHA user. The objective of the agent is to achieve proportional fairness.}
		\label{fig:eva2_1}
	\end{figure}
	
	Fig. \ref{fig:eva1} presents the short-term sum throughput---the short-term throughput is defined as the average successful packet rate in the past $W=2000$ time slots---of the agent, TDMA and ALOHA. In each subfigure of Fig. \ref{fig:eva1}, we fix the value of  $e_{down}$, and change the value of  $K$ (i.e., the history length of the transmission results maintained in the ACK packet) in the feedback recovery mechanism. The black line in Fig. \ref{fig:eva1} is the sum throughput benchmark. Each line except the black line is the average result over 10 different experiments with the shaded areas being areas within the standard deviation.
	
	As we can see from Fig. \ref{fig:eva1}, for  $e_{down} = 0.1, 0.2, 0.4, 0.6$, the results corresponding to $K = 1$  all fail to approximate the benchmark.  In addition, when $e_{down}$  increases, the gap between the results of  $K = 1$ and the benchmark also increases. This is because when  $K = 1$, no incomplete experiences are made complete and all the incomplete experiences are discarded. Without having enough complete experiences, especially when $e_{down}$   is large, the agent may not learn a good solution. 
	
	We can also see that when  $e_{down}$  is small, e.g.,  $e_{down} = 0.1$, the results corresponding to  $K = 2, 8, 16$ all approximate the benchmark. As  $e_{down}$  increases, the result corresponding to $K = 2$  gets worse and fails to approximate the benchmark when  $e_{down} = 0.6$ (the value of  $e_{down}$   could be as large as 0.6 when the AP is blocked by some buildings in a real wireless communication scenario). The reason for this phenomenon is that in our feedback recovery mechanism, the probability of an incomplete experience without being made complete is  $e_{down}^{K}$    (as mentioned in Section \ref{sec:neural_net_train}). When   $e_{down} = 0.6$  and  $K = 2$,  $e_{down}^{K} = 0.36$. Therefore, around 36\% experiences are deemed to be incomplete, resulting in severe performance degradation. We can conclude that a small value of $K$ is not a good option. However, when  $K$ increases, the agent checks more incomplete experiences whenever the ACK is received, which may need more computation each time. Unless stated otherwise, henceforth we adopt a modest value of  $K = 8$  in our evaluation.

	We then study the performance of DLMA in reducing the detrimental effect of the imperfect feedback channels when the DLMA agent aims to achieve a different objective. Here, to conserve space, we only present results when the objective is to achieve proportional fairness (i.e.,  $
	\alpha = 1$). Fig. \ref{fig:eva2} presents the short-term sum-log throughput---the sum-log throughput corresponds to  $
	\alpha = 1$  in \eqref{eqn:fairness}---of the agent, TDMA and ALOHA. As we can see from Fig. \ref{fig:eva2}, the same observation can be found and the same conclusion can be made as in Fig. \ref{fig:eva1}. Overall, both Fig. \ref{fig:eva1} and \ref{fig:eva2} demonstrate the capability of our feedback recovery mechanism in reducing the detrimental effect of the imperfect feedback channels.

    We next evaluate the performance of DLMA when both the uplink channels and the downlink channels are imperfect (i.e., $e_{up} > 0$ and $e_{down} > 0$). Let us first consider the same setting  as in Fig. \ref{fig:eva1} with $K=8$. Fig. \ref{fig:eva1_1} presents the sum throughput of the DLMA agent, TDMA and ALOHA when different values of $e_{up}$  and  $e_{down}$  are considered. In addition, the corresponding benchmarks are also plotted in Fig. \ref{fig:eva1_1}. Note that due to the inherent and unavoidable uplink channel errors, the benchmarks for different values of  $e_{up}$  are also different---when  $e_{up}$ increases, the sum throughput benchmark decreases \cite{benchmark}. As we can see from Fig. \ref{fig:eva1_1}, the performance of DLMA can approach the benchmark performance in different imperfect channel scenarios. When we change the objective of the agent in Fig. \ref{fig:eva1_1} to achieving proportional fairness, as we can see from Fig. \ref{fig:eva2_1}, the sum-log throughput benchmarks can also be achieved in different imperfect channel scenarios. This demonstrates that our DLMA algorithm can also work well when the uplink channels are imperfect.

	\subsubsection{Multiple Agents with Perfect Channels}\label{sec:eva2}
	This subsection examines the capability of the two-stage action selection approach to tackle the multi-agent problem. Specifically, we consider a case where four DLMA agents coexist with one TDMA user. As in Section \ref{sec:eva1}), TDMA transmits in the \nth{2} slot out of 5 slots within a TDMA frame.  In this part, we assume the channels are perfect (i.e., $e_{up} = 0$  and  $e_{down} = 0$) and we set $K$  to 1 since there is no need to perform feedback recovery when the channels are perfect. We vary the value of  $\alpha$ in our evaluation. The benchmark for different values of  $\alpha$  is the same and is given below: the throughput of each agent is 0.2, and the throughput of TDMA is also 0.2.

	Table \ref{table:eva2-1} summarizes the individual throughputs of all the agents and the TDMA user when the agents adopt different values of  $\alpha$. As we can see from Table \ref{table:eva2-1}, the throughputs of all the users approximate the optimal result 0.2. This demonstrates that these four agents not only learn the transmission pattern of TDMA but also avoid collisions with each other when the channels are perfect.  
	
	\begin{table}[!t]
		\centering
		\caption{Individual throughputs of each agent and TDMA when the channels are perfect.}
		\label{table:eva2-1}
		\renewcommand\arraystretch{1.2}
		\begin{tabular}{|l|l|l|l|l|l|}
			\cline{1-6}
			& \textbf{Agent 1} & \textbf{Agent 1} & \textbf{Agent 1} & \textbf{Agent 1} & \textbf{TDMA}  \\ \cline{1-6}
			$\alpha = 0$ & 0.1995 & 0.1995 & 0.19995 & 0.1995 & 0.1971\\ \cline{1-6}
			$\alpha = 0.5$ & 0.1985 & 0.1985 & 0.1985 & 0.1985 & 0.1925\\ \cline{1-6}
			$\alpha = 1$ & 0.1985 & 0.1985 & 0.1985 & 0.1985 & 0.1967\\ \cline{1-6}
			$\alpha = 10$ & 0.1985 & 0.1985 & 0.1985 & 0.1985 & 0.1974\\ \cline{1-6}
			$\alpha = 50$ & 0.1985 & 0.1985 & 0.1985 & 0.1985 & 0.1985\\ \cline{1-6}
		\end{tabular}
	\end{table}
	
	\begin{table}[!t]
		\centering
		\caption{Individual throughputs of each agent and TDMA when the channels are  imperfect and dependent.}
		\label{table:eva2-2}
		\renewcommand\arraystretch{1.2}
		\begin{tabular}{|l|l|l|l|l|l|}
			\cline{1-6}
			& \textbf{Agent 1} & \textbf{Agent 1} & \textbf{Agent 1} & \textbf{Agent 1} & \textbf{TDMA}  \\ \cline{1-6}
			$\alpha = 0$ & 0.1938 & 0.1938 & 0.1938 & 0.1938 & 0.1921\\ \cline{1-6}
			$\alpha = 0.5$ & 0.1947 & 0.1947 & 0.1947 & 0.1947 & 0.1935\\ \cline{1-6}
			$\alpha = 1$ & 0.198 & 0.198 & 0.198 & 0.198 & 0.198\\ \cline{1-6}
			$\alpha = 10$ & 0.1973 & 0.1973 & 0.1973 & 0.1973 & 0.1939\\ \cline{1-6}
			$\alpha = 50$ & 0.19 & 0.19 & 0.19 & 0.19 & 0.19\\ \cline{1-6}
		\end{tabular}
	\end{table}
	
	\begin{table}[!t]
		\centering
		\caption{Individual throughputs of each agent and TDMA when the channels are  imperfect and independent.}
		\label{table:eva2-3}
		\renewcommand\arraystretch{1.2}
		\begin{tabular}{|l|l|l|l|l|l|}
			\cline{1-6}
			& \textbf{Agent 1} & \textbf{Agent 1} & \textbf{Agent 1} & \textbf{Agent 1} & \textbf{TDMA}  \\ \cline{1-6}
			$\alpha = 0$ & 0.1839 & 0.1839 & 0.1839 & 0.1839 & 0.1978\\ \cline{1-6}
			$\alpha = 0.5$ & 0.1837 & 0.1837 & 0.1837 & 0.1837 & 0.198\\ \cline{1-6}
			$\alpha = 1$ & 0.1837 & 0.1837 & 0.1837 & 0.1837 & 0.1938\\ \cline{1-6}
			$\alpha = 10$ & 0.1842 & 0.1842 & 0.1842 & 0.1842 & 0.1931\\ \cline{1-6}
			$\alpha = 50$ & 0.182 & 0.182 & 0.182 & 0.182 & 0.1923 \\ \cline{1-6}
		\end{tabular}
	\end{table}
	
	\subsubsection{Multiple Agents with Imperfect Channels}
	This subsection studies the multi-agent scenario with imperfect feedback channels. We first consider the coexistence of four agents with one TDMA user. The setups are the same as in Section \ref{sec:eva2}) except that we set  $e_{down} = 0.1$ here. As in Section \ref{sec:eva2}, the optimal benchmark used for each user is also 0.2. For this coexistence case, we consider two subcases: (i) the feedback channels between the agents and the AP are completely dependent: specifically, if a channel error happens (does not happen), it happens (does not happen) to all the channels at the same time\footnote{This will be the case when there are obstacles between the AP and the users, and the obstacles are very close to the AP. For example, if the AP is very close to a train station, when there are no trains in the station, the channels between the AP and the users will be in good condition; when there is a train or more than one trains in the station, the channels may be blocked simultaneously.}; (ii) the feedback channels are independent, i.e., the channel error happens independently on each channel. 
	
	Table \ref{table:eva2-2} and Table \ref{table:eva2-3} summarize the individual throughputs of all the agents and the TDMA user for subcase (i) and (ii), respectively. As we can see from Table \ref{table:eva2-2}, the throughput of each user approximates the optimal result 0.2 for different values of  $\alpha$. This demonstrates that when the channels are imperfect and dependent, the agents not only learn the transmission pattern of TDMA but also learn to avoid collisions with other agents. 
	
	\begin{figure}[!t]
		\centering
		\includegraphics[scale=0.43]{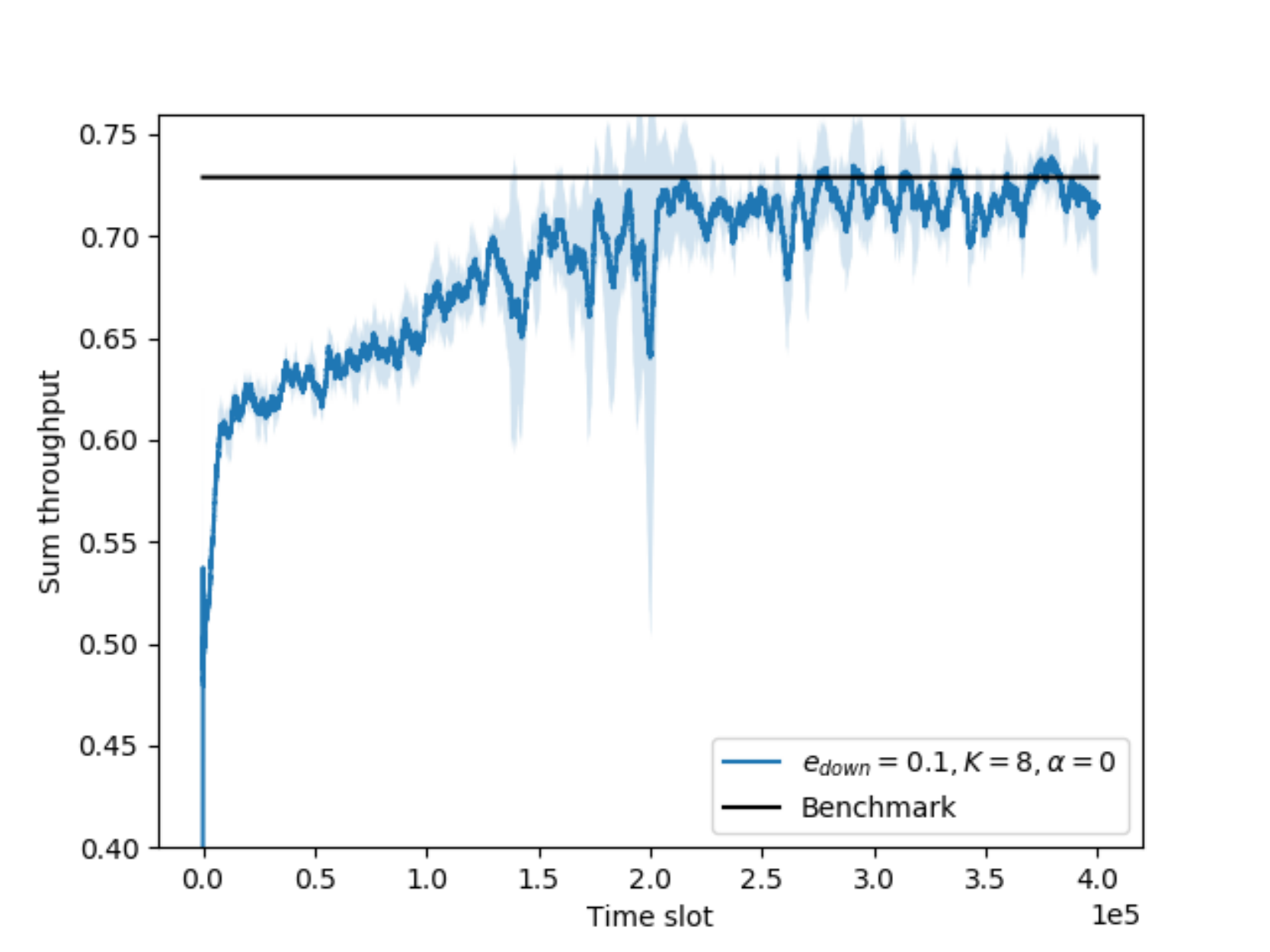}
		\caption[]{Sum throughput when five agents coexist with two TDMA users and three ALOHA users. The objective of the agents is to maximize sum throughput.}
		\label{fig:eva3}
	\end{figure}
	
	\begin{figure}[!t]
		\centering
		\includegraphics[scale=0.43]{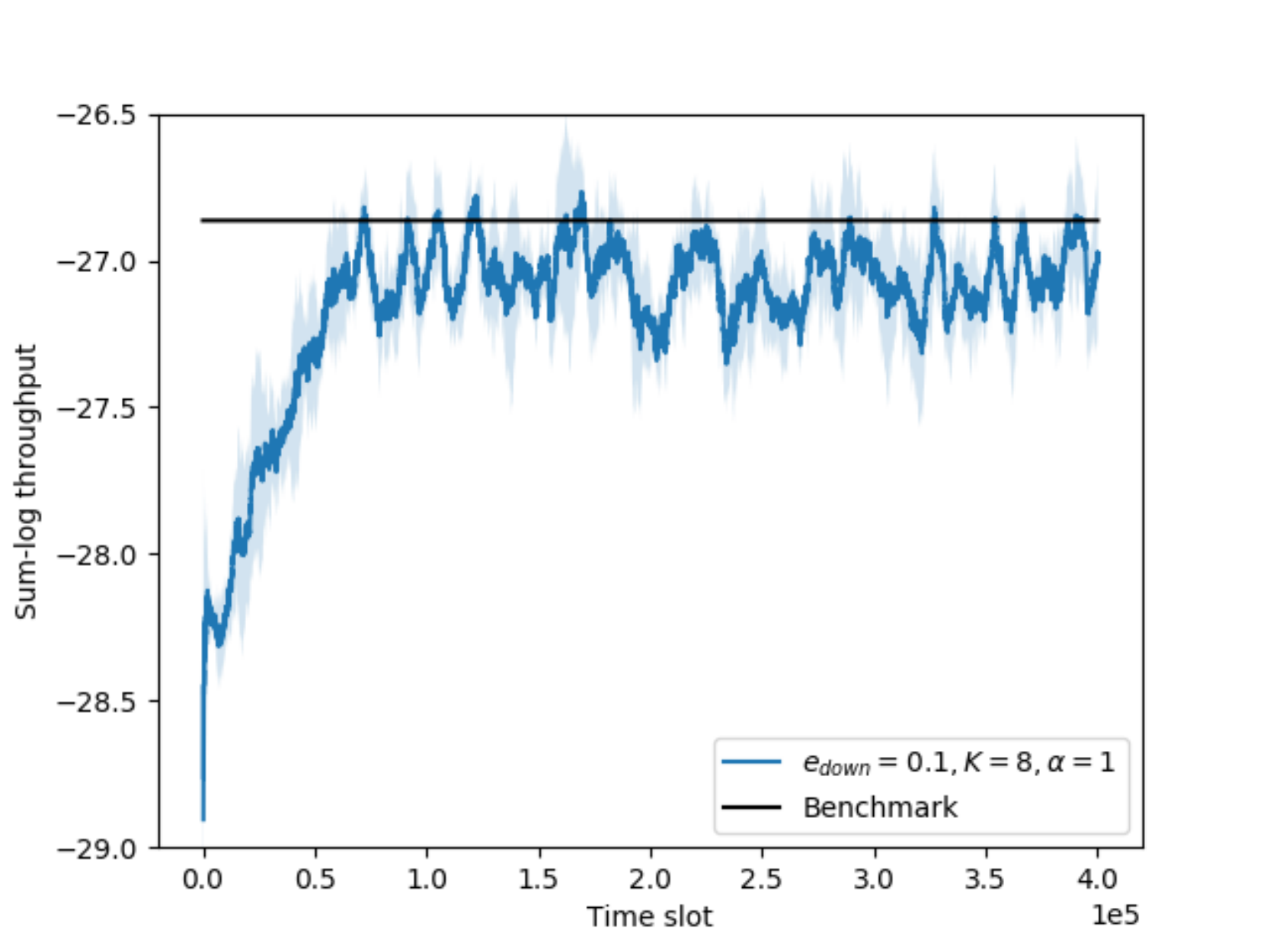}
		\caption[]{Sum-log throughput when five agents coexist with two TDMA users and three ALOHA users. The objective of the agents is to achieve proportional fairness.}
		\label{fig:eva4}
	\end{figure}

	As we can see from Table \ref{table:eva2-3}, only the throughput of TDMA approximates the optimal result 0.2, while the throughput of each agent is only around 0.18, which has a gap between the optimal result.  The reason for the gap is as follows. In subcase (i), the throughput vector that is used to break ties among the agents (see Section \ref{sec:two_stage}) known to each agent in each time slot is the same. While in subcase (ii), the throughput vector known to each agent may be different in each time slot since some agents may receive the ACK correctly and some may not. The discrepancy in the throughput vectors may cause more than one agent to transmit in each time slot, resulting in an inter-agent collision. Furthermore, for subcase (ii), we can analyze that the probability of the four agents having the same throughput vector in each time slot at least is ${\left( {1 - {e_{down}}} \right)^4} = 0.6561$. Therefore, the lower bound of each agent's throughput should be  $0.6561 \times 0.2 = 0.13122$. We can see that the throughput of each agent in Table \ref{table:eva2-3} falls in the range of  $\left[ {0.13122,0.2} \right]$. Therefore, we can conclude that when the channels are imperfect and independent, the agents can learn the transmission pattern of TDMA and reduce, rather than avoid, the collisions among the agents. 
	
	We next consider a more complex scenario where five DLMA agents coexist with two TDMA users and three ALOHA users. The first TDMA user transmits in the \nth{2} slot out of 10 slots within a TDMA frame, and the second TDMA user transmits in the \nth{8} slot out of 10 slots within a TDMA frame. The transmission probability of each ALOHA user is  0.1. The uplink channel error probability is  $e_{up} = 0$ and the downlink channel error probability is  $e_{down} = 0.1$.  For simplicity, here we assume the channels between the agents and the DLMA AP are dependent. Fig. \ref{fig:eva3} and Fig. \ref{fig:eva4} present the sum throughput and the sum-log throughput of all the users when the objectives of the agents are to maximize sum throughput and to achieve proportional fairness, respectively. In addition, the corresponding benchmarks are also plotted in Fig. \ref{fig:eva3} and Fig. \ref{fig:eva4}. As we can see from Fig. \ref{fig:eva3} and Fig. \ref{fig:eva4}, the optimal benchmarks can also be approximated for this complex scenario.

	\section{Conclusion}
	
	This paper developed a distributed DLMA protocol for efficient and equitable spectrum sharing in heterogeneous wireless networks with imperfect channels. Each user in the DLMA network is regarded as an agent and adopts the DLMA protocol to coexist with other DLMA agents and the users from the other networks. The agents do not have collaboration with each other and are unaware of the operational details of the MACs of  the users of other networks. Through a DRL process, the agents learn to work together to achieve a global  $\alpha$-fairness objective. 
	
	The conventional RL/DRL framework assumes that the feedback/reward to the agent is always correctly received. In wireless networks, however, the feedback/reward may be lost due to channel errors. Without correct feedback information, the agent may fail to find a good solution. Moreover, in the distributed protocol, each agent makes decisions on its own. It is a challenge to guarantee that the multiple agents will make coherent decisions and work together to achieve the same objective, particularly in the face of imperfect feedback channels. To deal with the challenge, we put forth a feedback recovery mechanism to recover missing feedback information and  a two-stage action selection mechanism to aid coherent decision making to reduce transmission collisions among the agents. Extensive simulation results demonstrated the effectiveness of these two mechanisms. 
	
	Last but not least, we believe that the feedback recovery mechanism and the two-stage action selection mechanism proposed in this paper can also be used in general distributed multi-agent reinforcement learning problems in which feedback information on rewards can be corrupted.

	\ifCLASSOPTIONcaptionsoff
	\newpage
	\fi

	{\footnotesize 
		\bibliographystyle{IEEEtran}
		\bibliography{DLMA_plus}

\begin{thebibliography}{10}
\providecommand{\url}[1]{#1}
\csname url@samestyle\endcsname
\providecommand{\newblock}{\relax}
\providecommand{\bibinfo}[2]{#2}
\providecommand{\BIBentrySTDinterwordspacing}{\spaceskip=0pt\relax}
\providecommand{\BIBentryALTinterwordstretchfactor}{4}
\providecommand{\BIBentryALTinterwordspacing}{\spaceskip=\fontdimen2\font plus
\BIBentryALTinterwordstretchfactor\fontdimen3\font minus
  \fontdimen4\font\relax}
\providecommand{\BIBforeignlanguage}[2]{{%
\expandafter\ifx\csname l@#1\endcsname\relax
\typeout{** WARNING: IEEEtran.bst: No hyphenation pattern has been}%
\typeout{** loaded for the language `#1'. Using the pattern for}%
\typeout{** the default language instead.}%
\else
\language=\csname l@#1\endcsname
\fi
#2}}
\providecommand{\BIBdecl}{\relax}
\BIBdecl

\bibitem{tilghman2019will}
P.~Tilghman, ``Will rule the airwaves: A darpa grand challenge seeks autonomous
  radios to manage the wireless spectrum,'' \emph{IEEE Spectrum}, vol.~56,
  no.~6, pp. 28--33, 2019.

\bibitem{liang2008sensing}
Y.-C. Liang, Y.~Zeng, E.~C. Peh, and A.~T. Hoang, ``Sensing-throughput tradeoff
  for cognitive radio networks,'' \emph{IEEE transactions on Wireless
  Communications}, vol.~7, no.~4, pp. 1326--1337, 2008.

\bibitem{sutton2018reinforcement}
R.~S. Sutton and A.~G. Barto, \emph{Reinforcement learning: An
  introduction}.\hskip 1em plus 0.5em minus 0.4em\relax MIT press, 2018.

\bibitem{lecun2015deep}
I.~Goodfellow, Y.~Bengio, and A.~Courville, \emph{Deep learning}.\hskip 1em
  plus 0.5em minus 0.4em\relax MIT press, 2016.

\bibitem{DQNpaper}
V.~Mnih, K.~Kavukcuoglu, D.~Silver, A.~A. Rusu, J.~Veness, M.~G. Bellemare,
  A.~Graves, M.~Riedmiller, A.~K. Fidjeland, G.~Ostrovski \emph{et~al.},
  ``Human-level control through deep reinforcement learning,'' \emph{Nature},
  vol. 518, no. 7540, pp. 529--533, 2015.

\bibitem{silver2016mastering}
D.~Silver, A.~Huang, C.~J. Maddison, A.~Guez, L.~Sifre, G.~Van Den~Driessche,
  J.~Schrittwieser, I.~Antonoglou, V.~Panneershelvam, M.~Lanctot \emph{et~al.},
  ``Mastering the game of go with deep neural networks and tree search,''
  \emph{nature}, vol. 529, no. 7587, p. 484, 2016.

\bibitem{gu2017deep}
S.~Gu, E.~Holly, T.~Lillicrap, and S.~Levine, ``Deep reinforcement learning for
  robotic manipulation with asynchronous off-policy updates,'' in \emph{2017
  IEEE international conference on robotics and automation (ICRA)}.\hskip 1em
  plus 0.5em minus 0.4em\relax IEEE, 2017, pp. 3389--3396.

\bibitem{luong2019applications}
N.~C. Luong, D.~T. Hoang, S.~Gong, D.~Niyato, P.~Wang, Y.-C. Liang, and D.~I.
  Kim, ``Applications of deep reinforcement learning in communications and
  networking: A survey,'' \emph{IEEE Communications Surveys \& Tutorials},
  2019.

\bibitem{sun2019application}
Y.~Sun, M.~Peng, Y.~Zhou, Y.~Huang, and S.~Mao, ``Application of machine
  learning in wireless networks: Key techniques and open issues,'' \emph{IEEE
  Communications Surveys \& Tutorials}, 2019.

\bibitem{shao2019significant}
Y.~Shao, A.~Rezaee, S.~C. Liew, and V.~W. Chan, ``Significant sampling for
  shortest path routing: A deep reinforcement learning solution,'' in
  \emph{2019 IEEE Global Communications Conference (GLOBECOM)}.\hskip 1em plus
  0.5em minus 0.4em\relax IEEE, 2019, pp. 1--7.

\bibitem{mo2000fair}
J.~Mo and J.~Walrand, ``Fair end-to-end window-based congestion control,''
  \emph{IEEE/ACM Transactions on Networking (ToN)}, vol.~8, no.~5, pp.
  556--567, 2000.

\bibitem{yu2019deep}
Y.~Yu, T.~Wang, and S.~C. Liew, ``Deep-reinforcement learning multiple access
  for heterogeneous wireless networks,'' \emph{IEEE Journal on Selected Areas
  in Communications}, vol.~37, no.~6, pp. 1277--1290, 2019.

\bibitem{zhang2019multi}
K.~Zhang, Z.~Yang, and T.~Ba{\c{s}}ar, ``Multi-agent reinforcement learning: A
  selective overview of theories and algorithms,'' \emph{arXiv preprint
  arXiv:1911.10635}, 2019.

\bibitem{ijcai2017-656}
\BIBentryALTinterwordspacing
T.~Everitt, V.~Krakovna, L.~Orseau, and S.~Legg, ``Reinforcement learning with
  a corrupted reward channel,'' in \emph{Proceedings of the Twenty-Sixth
  International Joint Conference on Artificial Intelligence, {IJCAI-17}}, 2017,
  pp. 4705--4713. [Online]. Available:
  \url{https://doi.org/10.24963/ijcai.2017/656}
\BIBentrySTDinterwordspacing

\bibitem{wang2020rlnoisy}
J.~Wang, Y.~Liu, and B.~Li, ``Reinforcement learning with perturbed rewards,''
  in \emph{AAAI}, 2020.

\bibitem{naparstek2018deep}
O.~Naparstek and K.~Cohen, ``Deep multi-user reinforcement learning for
  distributed dynamic spectrum access,'' \emph{IEEE Transactions on Wireless
  Communications}, vol.~18, no.~1, pp. 310--323, 2018.

\bibitem{wang2018deep}
S.~Wang, H.~Liu, P.~H. Gomes, and B.~Krishnamachari, ``Deep reinforcement
  learning for dynamic multichannel access in wireless networks,'' \emph{IEEE
  Transactions on Cognitive Communications and Networking}, vol.~4, no.~2, pp.
  257--265, 2018.

\bibitem{chang2018distributive}
H.-H. Chang, H.~Song, Y.~Yi, J.~Zhang, H.~He, and L.~Liu, ``Distributive
  dynamic spectrum access through deep reinforcement learning: A reservoir
  computing-based approach,'' \emph{IEEE Internet of Things Journal}, vol.~6,
  no.~2, pp. 1938--1948, 2018.

\bibitem{zhong2018actor}
C.~Zhong, Z.~Lu, M.~C. Gursoy, and S.~Velipasalar, ``Actor-critic deep
  reinforcement learning for dynamic multichannel access,'' in \emph{2018 IEEE
  Global Conference on Signal and Information Processing (GlobalSIP)}.\hskip
  1em plus 0.5em minus 0.4em\relax IEEE, 2018, pp. 599--603.

\bibitem{xu2018deep}
Y.~Xu, J.~Yu, W.~C. Headley, and R.~M. Buehrer, ``Deep reinforcement learning
  for dynamic spectrum access in wireless networks,'' in \emph{MILCOM 2018-2018
  IEEE Military Communications Conference (MILCOM)}.\hskip 1em plus 0.5em minus
  0.4em\relax IEEE, 2018, pp. 207--212.

\bibitem{tan2019deep}
J.~Tan, L.~Zhang, Y.-C. Liang, and D.~Niyato, ``Deep reinforcement learning for
  the coexistence of laa-lte and wifi systems,'' in \emph{ICC 2019-2019 IEEE
  International Conference on Communications (ICC)}.\hskip 1em plus 0.5em minus
  0.4em\relax IEEE, 2019, pp. 1--6.

\bibitem{watkins1992q}
C.~J. Watkins and P.~Dayan, ``Q-learning,'' \emph{Machine learning}, vol.~8,
  no. 3-4, pp. 279--292, 1992.

\bibitem{lin1992self}
L.-J. Lin, ``Self-improving reactive agents based on reinforcement learning,
  planning and teaching,'' \emph{Machine learning}, vol.~8, no. 3-4, pp.
  293--321, 1992.

\bibitem{bertsekas1992data}
D.~P. Bertsekas, R.~G. Gallager, and P.~Humblet, \emph{Data networks}.\hskip
  1em plus 0.5em minus 0.4em\relax Prentice-Hall International New Jersey,
  1992, vol.~2.

\bibitem{hochreiter1997long}
S.~Hochreiter and J.~Schmidhuber, ``Long short-term memory,'' \emph{Neural
  computation}, vol.~9, no.~8, pp. 1735--1780, 1997.

\bibitem{benchmark}
Y.~Yu, T.~Wang, and S.~C. Liew, ``Multi-agent deep reinforcement learning
  multiple access for heterogeneous wireless networks with imperfect
  channels,'' \emph{Benchmark, \textnormal{available at:}
  \textnormal{\url{https://github.com/YidingYu/DLMA/blob/master/MA-DLMA-benchmark.pdf}}}.

\end{thebibliography}
	}
\end{document}